\documentclass[preprint]{elsarticle}
\usepackage{amsmath,amssymb,color}
\usepackage{algorithm,algpseudocode,rotating}
\usepackage{lineno,hyperref}
\usepackage{pdflscape,color}
\usepackage{graphicx, eurosym, indentfirst,float,latexsym}
\usepackage{amsmath,amsthm,amsopn,amstext,amscd,amsfonts,amssymb}
\modulolinenumbers[0]
\biboptions{authoryear}
\journal{BioSystems}

\DeclareMathOperator{\asinh}{arcsinh}
\newcommand{\x}{x_{ij}}
\newcommand{\y}{x_{i,j-1}}
\newcommand{\vt}{\Delta_{ij}} 
\newcommand{\ds}{\sum_{i=1}^{d}\sum_{j=2}^{n_i}}
\renewcommand{\l}{R_{ij}} 

\renewcommand{\S}{S_{ij}^{\eta,\lambda,\mu}}
\newcommand{\T}{T_{ij}^{\eta,\lambda,\mu}}
\newcommand{\V}{V_{ij}^{\eta,\lambda,\mu}}
\newcommand{\W}{W_{ij}^{\eta,\lambda,\mu}}
\newcommand{\Z}{Z_{ij}^{\eta,\lambda,\mu}}
\newcommand{\para}{{\eta,\lambda,\mu}}

\begin{document}

\begin{frontmatter}

\title{A hyperbolastic type-I diffusion process: parameter estimation by means of the firefly algorithm}


\author[barrera]{Antonio Barrera}
\ead{antonio.barrera@uma.es}

\author[ugr]{Patricia Rom\'an-Rom\'an}
\ead{proman@ugr.es}

\author[ugr]{Francisco Torres-Ruiz\corref{corr}}
\ead{fdeasis@ugr.es}

\cortext[corr]{Corresponding author}

\address[barrera]{Departamento de Matem\'atica Aplicada, E.T.S.I. Inform\'atica, Bulevar Louis Pasteur, 35, Campus de Teatinos, Universidad de M\'alaga, 29071 M\'alaga, Spain.}

\address[ugr]{Departamento de Estad\'istica e Investigaci\'on Operativa, Facultad de Ciencias, s/n, Campus de Fuentenueva, Universidad de Granada, 18071 Granada, Spain.}
\date{}

\begin{abstract}
  A stochastic diffusion process, whose mean function is a hyperbolastic curve of type I, is presented. The main characteristics of the process are studied and the problem of maximum likelihood estimation for the parameters of the process is considered. To this end, the firefly metaheuristic optimization algorithm is applied after bounding the parametric space by a stagewise procedure. Some examples based on simulated sample paths and real data illustrate this development.
\end{abstract}


\end{frontmatter}


\section{Introduction}

The construction of mathematical models to describe growth dynamics has been the subject of several studies in the last decades. The reason is the variety of situations where these phenomena arise in a natural way. Originally, studies focused on population growth, although nowadays they extend to many other research fields. For instance, in recent years, these models have been applied to tumor growth and the spread of diseases.

One of the main problems associated with the study of growth phenomena is the choice of a suitable model: even though diverse representations by deterministic models based on sigmoidal curves have been used (\cite{tsoularis_2002}), the presence of fixed inflection points restricts the adequacy of the model to real circumstances.

In order to deal with these issues, the hyperbolastic curves of type I, II and III (H1, H2 and H3, respectively), were developed by \cite{tabatabai_2005}, who introduced hyperbolic functions into known models, thus obtaining mobile inflection points and increasing the capability of the models to fit real data.

Recent results have proved the usefulness of these curves in the description and modeling of dynamical phenomena. In this sense, \cite{eby_2010} used hyperbolastic models to study the growth of the solid Ehrlich carcinoma under particular treatments, obtaining a more accurate representation than those yielded by other classic curves such as Gompertz or Weibull. \cite{tabatabai_2011} used the H3 model to describe the behavior of embryonic stem cells, improving the results of other models such as those in \cite{deasy2003}. Recently, new models also derived from hyperbolastic curves have been introduced, such as the oscillabolastic model (\cite{tabatabai_2012}) or the T-model (\cite{tabatabai_2013_x}).

Despite the good results obtained by applying these deterministic models, it is clear that natural dynamical phenomena occur under the influence of unknown (or even immeasurable) factors which can have a significant effect on the evolution of the process. Therefore, introducing into the models certain elements capable of describing such influences seems necessary and justified. In this way, the next step in the mathematical modeling of dynamic phenomena lies in extending deterministic models to stochastic models while taking into account the random influences affecting the dynamics of the phenomenon.

To this end, stochastic processes, and in particular diffusion processes, appear as the most appropriate tool. In the context of growth curves, these processes arise when a random factor is introduced into the differential equation whose solution is the aforementioned curve, thus becoming a stochastic differential equation whose solution is the final process.

This methodology, used by \cite{capocelliricciardi_1974} for obtaining a diffusion process associated with the Gompertz curve, has been applied to a wide range of curves. A modification of such model has been recently applied to tumor growth by \cite{albano_2011, Albano2013}.
Diffusion processes based on the Bertalanffy curve have been applied to the study of population growth in animals and plants (\cite{roman_2010}; \cite{roman_2014}; \cite{russo_2009}). With respect to the logistic curve, for which the first model was proposed by \cite{Feller1940}, Kolmogorov equations have no explicit solution, so the transition probability density function is not known in closed-form. Therefore, many variants have been proposed, for which a summary can be consulted in \cite{tuckwell_1987}. \cite{schurz_2007} considered a more general version of the stochastic differential equation associated with logistic growth and \cite{Barrera-Garcia2013a} proposed a Gaussian logistic diffusion process, whereas \cite{roman_2012} established a diffusion process based on a reformulation of the logistic curve. Recently, \cite{roman_2015} generalized this last diffusion process to the case of the Richards curve.

In this paper we introduce a diffusion process whose mean function is a reformulation of the H1 curve. This allows us to express the asymptotic behavior according to the initial values. Thanks to this, we can consider situations where data are available from many individuals, each exhibiting the same growth pattern but with different bounds for the initial value.

In this way, we construct a model such that its mean function is an H1 curve, making it suitable for prediction purposes. To this end, data must be used to obtain estimates for the parameters of the process, an estimation that is carried out by maximum likelihood. This is not problematic as far the parameters of the initial distribution are concerned, but the estimation of the rest of parameters yields a system of equations whose solution is not guaranteed by classical numerical procedures. An alternative is the use of metaheuristic optimization procedures, to which purpose we propose using the firefly algorithm (FA). Introduced by \cite{Yang2008}, this population-based, nature-inspired metaheuristic algorithm is undergoing an important development in fields such as engineering and optimization (e.g. the works of \cite{alb2016} and \cite{kavousi_2014} for its efficiency and capacity to deal with NP-hard problems). Some interesting modifications have also been made (\cite{gandomi_2013}; \cite{zang_2016}). The
low computational cost of the algorithm makes it especially useful to address maximum likelihood estimation in diffusion processes.

This paper is structured as follows: Section 2 presents a reformulation of the H1 curve, including some relevant properties. The scope of this work is restricted to the case of increasing curves showing at least one inflection point. The hyperbolastic type-I diffusion process (from now on referred to as H1 diffusion process) is introduced as a particular case of the lognormal diffusion process with exogenous factors. Section 3 deals with the estimation of the parameters of the model by means of the maximum likelihood procedure, showing the complexity of the system of equations. This question leads to our proposal of using FA as a valid tool for maximizing the likelihood function. A summary of the algorithm and its properties is then shown, as well as the modifications required in our context. In this sense, a procedure for bounding the parametric space is proposed. Finally, in Section 4, the methodology is applied to two examples: the first over simulated data, and the second over real data from a study
about a
molecular biology technique called quantitative polymerase chain reaction.

\section{The model}

\subsection{Reformulation of the H1 curve} \label{subsecparam}

The H1 curve is the solution of the Bernoulli differential equation
\begin{equation}\label{pode}
\frac{dx(t)}{dt} = M^{-1}x(t) \Big(M - x(t)\Big) \left(\rho + \frac{\theta}{ \sqrt{1+t^2}} \right), \, t\geq t_0, \, \theta \in \mathbb{R},
\end{equation}
with initial value $x(t_0)=x_0>0$. Here, $x(t)$ represents the size of a population at time instant $t$ and $M$ denotes the maximum sustainable population (carrying capacity), whereas parameters $\theta$ and $\rho$ jointly determine growth rate. Note that when $\theta = 0$ this equation is the well-known logistic differential equation.

The solution to (\ref{pode}) is
\begin{equation}\label{p}
x(t) = M \left(1 + a\, \exp \left( - \rho t - \theta\asinh(t) \right) \right)^{-1},
\end{equation}
where
\begin{equation*}
a= x_0^{-1}(M-x_0)\,\exp \left( \rho t_0 + \theta\asinh(t_0) \right).
\end{equation*}

Note that $\lim_{t\rightarrow \infty}x(t)=M$ if $\rho>0$, and $\lim_{t\rightarrow \infty}x(t)=0$ if $\rho<0$. So, if $\rho>0$, the curve increases to the asymptote $M$, whereas $\rho<0$ leads to decay profiles.

In the above remark we have shown that the asymptote $M$ is independent from the initial value, something that can become an important restriction for some applications. In practice, there are situations in which the growth phenomenon under study shows an H1-type sigmoidal growth and several sample paths are available, each with the same growth pattern but with different initial values and upper bounds (as the weight of individuals in a population). For these reasons, in order to model the situations in which the limit value depends on the initial one, we consider a reformulation of curve (\ref{p}). Setting $\eta=1/a$, $\lambda=e^{-\rho}$, and $\mu=e^{-\theta}$ we obtain
\begin{equation}\label{H1}
x(t) = x_0\frac{\eta + \xi(t_0)}{\eta + \xi(t)}, t\geq t_0, \, \eta,\,\mu>0,
\end{equation}
where $\xi(t)=\lambda^{t} \mu^{\asinh{(t)}}$.
In the following we will deal with increasing curves. This question leads to conditions
$0<\lambda<1$ and $\mu < c_{\lambda}(t)$, $\forall t\geq t_0$,  where $c_\lambda(t) = \lambda^{-\sqrt{1+t^2}}$ is an increasing function for all $t$. Therefore, in the following we will consider $\mu < c_\lambda(t_0)$.

Another important characteristic of the H1 curve is the mobility of its inflections, which gives it a flexibility superior to that of other commonly used curves, allowing for more precise modeling of phenomena showing a sigmoidal behavior.
From $x''(t)=0$ it follows that inflections of the curve verify the following equation, which corresponds, after the reparametrization made in the curve, with that obtained by Tabatabai et al. (2005):
\begin{equation*}\label{inflexion}
\frac{2\eta}{\eta+\xi(t)}-1 = \frac{t\,\log\mu}{(1+t^2)^{3/2}}\left(\log\lambda+\frac{\log\mu}{\sqrt{1+t^2}}\right)^{-2}.
\end{equation*}

\subsection{The H1 diffusion process}
\label{secdif}

In this section we will introduce a diffusion process associated with the curve (\ref{H1}). To this end, we will proceed with the general procedure of obtaining a stochastic differential equation from an ordinary differential equation associated with the curve. We will then verify that the mean function of the resulting process is an H1 curve, which is useful for the purpose of making predictions in situations modeled by this type of growth.

From (\ref{H1}) we can confirm that the curve verifies the differential equation
\[
 \frac{dx(t)}{dt} = h(t) x(t), \ \ \ x(t_0)=x_0,
\]
where
\begin{equation*}\label{h1}
h(t) = -\frac{d\xi(t)/dt}{\eta+\xi(t)},
\end{equation*}
which is a generalization of the Malthusian growth model with a time-dependent fertility rate $h(t)$.

Replacing this fertility rate with $h(t)+\sigma\,W(t)$, where $W(t)$ denotes the standard Wiener process and $\sigma>0$, we obtain the Langevin equation
\begin{equation*}
\frac{dX(t)}{dt}=h(t)X(t)+\sigma\,X(t)\,W(t),
\end{equation*}
which, rewritten as a stochastic differential equation leads to
\begin{equation}
\label{EDENuevo} dX(t)=h(t)X(t)dt+\sigma X(t)dW(t).
\end{equation}

Taking into account that $h$ is a continuous and bounded function, equation (\ref{EDENuevo}) verifies the conditions for the existence and uniqueness of solution (see \cite{oksendal_2003}). The solution is a stochastic diffusion process taking values on $\mathbb{R}^+$ and characterized by infinitesimal moments $A_1(x,t) = h(t)x$ and $A_2(x) = \sigma^2 x^2$. In addition, a closed-form expression for the solution can be provided. In fact, by considering the initial condition $X(t_0)=X_0$, independent from $W(t)$ for $t\geq t_0$, we have
\begin{equation}
\label{SOLEDE}
X(t)=X_0\frac{\eta+\xi(t_0)}{\eta+\xi(t)}\exp\left(-\frac{\sigma^2}{2}(t-t_0)+\sigma(W(t)-W(t_0))\right).
\end{equation}

This process is a particular case of the lognormal diffusion process with exogenous factors.
Several applied works have been developed around this process. For instance, \cite{gutierrez_1999, gutierrez_2003}, performed an inferential analysis and assessed its usefulness for studies in Economics, including the consideration of first-passage-time problems, a topic already considered in \cite{gutierrez_1995}.

As regards the distribution of the process, if $X_0$ is distributed according to a lognormal distribution $\Lambda_1\left[\mu_0;\sigma_0^2\right]$ or $X_0$ is a degenerate variable ($P[X_0=x_0]=1$),
all the finite-dimensional distributions of the process are lognormal. Concretely, $\forall n\in\mathbb{N}$ and $t_1<\ldots<t_n$, vector $(X(t_1),\ldots,X(t_n))^{T}$ has a $n$-dimensional lognormal distribution $\Lambda_n[\delta,\Sigma]$, where the components of vector $\delta$ and matrix $\Sigma$ are
$$
\delta_i=\mu_0+\log\left(\frac{\eta+\xi(t_0)}{\eta+\xi(t)}\right)-\frac{\sigma^2}{2}(t-t_0), \ \ i=1,\ldots,n
$$
and
$$
\sigma_{ij}=\sigma_0^2+\sigma^2(\min(t_i,t_j)-t_0), \ \ i,j=1,\ldots,n
$$
respectively. The transition probability density function can be obtained from the distribution of $(X(s),X(t))^T$, being
\begin{align}\label{dens}
f(x,t|,y,s) = & \frac{1}{x\sqrt{2\pi \sigma^2 (t-s)}} \exp{ \left(-\frac{\left(\log \frac{x}{y} - \log \frac{\eta + \xi(s)}{\eta + \xi(t)} + \frac{\sigma^2}{2}(t-s)\right)^2}{2\sigma^2 (t-s)}\right)} \nonumber
\end{align}
that is, $X(t) | X(s) = y$ follows a lognormal distribution
$$
X(t) | X(s) = y \rightsquigarrow \Lambda_1 \left( \log y + \log \frac{\eta + \xi(s) }{\eta + \xi(t) } - \frac{\sigma^2}{2}( t -s); \sigma^2(t-s) \right).
$$

From the previous distributions the main characteristics of the process can be found. If we note
$$
G_n(t|z,\tau)=z^n\left(\displaystyle\frac{\eta+\xi(\tau)}{\eta+\xi(t)}\right)^{n}
\exp\left(\frac{n(n-1)\sigma^2}{2}(t-\tau)\right), n\geq 0,
$$
then
\begin{equation*}
E\left[X(t)^n\right]=G_n\left(t|E[X(t_0)],t_0\right) \ \ \ \mbox{and} \ \ \ E\left[X(t)^n|X(\tau)=y\right]=G_n\left(t|y,\tau\right).
\end{equation*}

In particular, the mean function of the process and the one conditioned to an initial value $x_0$ are
\begin{equation*}
\label{Media1}
E\left[X(t)\right]=E[X(t_0)]\displaystyle\frac{\eta+\xi(t_0)}{\eta+\xi(t)}
\end{equation*}
and
\begin{equation*}
\label{Media2}
E\left[X(t)|X(t_0)=x_0\right]=x_0\displaystyle\frac{\eta+\xi(t_0)}{\eta+\xi(t)}
\end{equation*}
respectively, which are H1 curves of the type (\ref{H1}) described previously. This justifies considering the stochastic model herein proposed.

\section{Estimation of the model}
Let us consider a discrete sampling of the process, based on $d$ sample paths, for times $t_{ij}$, ($i=1,\ldots, d$, $j=1,\ldots, n_i)$ with $t_{i1}=t_1$, $i=1,\ldots,d$. That is, we observe variables $X(t_{ij})$, the values of which: $\mathbf{x}=\left\{x_{ij}\right\}_{i=1,\ldots,d; j=1,\ldots,n_i}$,
make up the sample for the inferential study.

By considering the most general case in which the initial distribution is lognormal, $X(t_{1})\sim\Lambda(\mu_1,\sigma_1^2)$, the log-likelihood function of the sample is

\begin{equation*}\label{ml}
\log L_{\mathbf{x}} (\mu_1,\sigma^2_1,\eta,\lambda,\mu,\sigma^2) =
\sum_{i=1}^{d}\log f_1(x_{i1}) + \sum_{i=1}^{d}\sum_{j=2}^{n_i} \log
f(x_{ij},t_{ij}|x_{i,j-1},t_{i,j-1})
\end{equation*}
where $f_1$ is the density function of $X(t_1)$.

Denoting $N=\displaystyle\sum_{i=1}^{d}n_i$, we have
\begin{eqnarray}
\label{verosimilitud}
    \log L_{\mathbf{x}}(\mu_1,\sigma_1^2,\eta,\lambda,\mu,\sigma^2)&=&
     -\frac{N}{2}\log(2\pi)-\frac d2\log\sigma_1^2-\frac{N-d}{2}\log\sigma^2-\sum_{i=1}^d\log x_{i1}\nonumber\\
     &-&\frac{1}{2\sigma_1^2}\sum_{i=1}^d\left[\log x_{i1}-\mu_1\right]^2-\sum_{i=1}^d\sum_{j=2}^{n_i}\log x_{ij}-\frac{1}{2}\sum_{i=1}^d\sum_{j=2}^{n_i}
        \log\vt\nonumber\\
      &-&\frac{1}{2\sigma^2}\ds\frac1{\vt}\left(\l - \T + \frac{\sigma^2}{2}\vt\right)^2
\end{eqnarray}
where
$$T_{ij}^{\eta,\lambda,\mu}=\log\left(\frac{\eta+\xi(t_{i,j-1})}{\eta+\xi(t_{ij})}\right), \ \
\l= \log{\frac{\x}{\y}}\, \mbox{and } \vt= t_{ij} - t_{i,j-1}.$$

From (\ref{verosimilitud}) the ML estimates of $\mu_1$ and $\sigma_1^2$ are
\[
  \widehat{\mu}_1=\displaystyle\frac{1}{d}\displaystyle\sum_{i=1}^{d}\log x_{i1} \ \ \mbox{and} \ \
  \widehat{\sigma}_1^2=\displaystyle\frac{1}{d}\displaystyle\sum_{i=1}^{d}(\log x_{i1}-\widehat{\mu}_1)^2.
\]

However, estimating the rest of the parameters poses some difficulties. Concretely, the resulting system of equations (see Appendix) is exceedingly complex and does not have an explicit solution. Therefore, numerical procedures must be employed to find its approximate solution.
Since the likelihood system of equations depends on sample data, it is impossible to carry out a general study about it. For instance, the conditions of convergence of the most widely used numerical methods are impossible to verify. Such methods (as Newton-Raphson and its variants) require calculating and inverting the Jacobian and Hessian matrices of the vectorial function that determines the system of equations. The sample data may then lead to singular matrices and, consequently, to a failure of the numerical procedure. Rom\'an et al (2012) contains an example of this (for a Gompertz-type diffusion process). Adding to this problem, we must also select an initial solution.

For these reasons, we propose the use of the local search metaheuristic firefly algorithm in order to maximize the likelihood function.
The study and development of metaheuristic optimization algorithms have grown considerably in the last years, with applications having been developed to several fields, including estimations in diffusion processes. For example, \cite{roman_2012} used simulated annealing for estimating the parameters of a Gompertz-type diffusion process, whereas \cite{roman_2015} suggested the variable neighborhood search method in combination with simulated annealing in the context of the Richards diffusion process.

Once we have found the estimates of $\mu_1$ and $\sigma_1^2$, the problem becomes maximizing function $\log L_{\mathbf{x}}(\widehat{\mu}_1,\widehat{\sigma}_1^2,\eta,\lambda,\mu,\sigma^2)$. So, from (\ref{verosimilitud}), the target function we will consider is
\begin{equation*}\label{funobj}
f_o(\lambda,\mu,\eta,\sigma^2) = - \frac{d(N-1)}{2} \log\sigma^2 - \frac{1}{2 \sigma^2} \ds\frac1{\vt}\left(\l - \T + \frac{\sigma^2}{2}\vt\right)^2.
\end{equation*}

\subsection{Application of the firefly algorithm}

FA is a stochastic, population-based, and nature-inspired metaheuristic algorithm developed by \cite{Yang2008}. It has been successfully applied in many fields such as optimization and engineering (see \cite{kavousi_2014,niknam_2012}) because of its efficiency and ability to deal with NP-hard problems, becoming one of the most important tools of swarm intelligence, a research subfield of artificial intelligence focused on the collective behavior of decentralized and self-organized systems. In the last years many works have explored in detail the theory of the firefly algorithm and its applications. See, for instance, \cite{fister_2013_2} and \cite{yang_2013_y}, as well as \cite{fister_2014} and \cite{gandomi_2013} where some modifications of the algorithm have been introduced.

The algorithm works over population subgroups, and thus it can deal efficiently with nonlinear and multimodal optimization problems. Indeed, FA can be viewed as a generalization of well-known algorithms such as particle swarm optimization (PSO), differential evolution, and simulated annealing (SA).

It is inspired by the flashing lights of fireflies, produced by a bioluminescence phenomenon and used for attraction purposes between them. Thus, in basic terms, $n$ fireflies are randomly distributed over the search space, with each of them having an associated light intensity that is dependent on its position. Then, at every generation, fireflies are attracted (attractiveness is proportional to light intensity) by the brighter neighbors according to the distance between them; that is, if a firefly flashes with high intensity, the closer ones will detect high attractiveness and fly towards it.

With the movement of fireflies, the attraction between them is updated to the new distances and light intensities, and the generation finishes by ranking the fireflies from the dimmest to the brightest one, which correspond to those that provide a lowest and highest value of the objective function, respectively. This procedure is repeated for a fixed number of generations, with each one reducing the randomness of fireflies' movements.

Looking at the matter in more detail, it is obvious that light intensity, and therefore attractiveness, varies according to the distance between fireflies and to the medium and its light-absorbing properties. The firefly algorithm uses this characteristics in order to search for the local and global maxima of a given function, defining the search space as the domain of the objective function and the light intensity as proportional to the value of the function at every point occupied by a firefly.
Therefore, measures of the attraction between fireflies must decrease when the distance and/or the absorption coefficient increase. Usually, this is formulated as
\begin{equation*}
\beta(r) = \beta_0 \exp \left( -\gamma r^2 \right),
\end{equation*}
where $\beta_0$ is the attractiveness at $r=0$ and $\gamma$ is the absorption coefficient of the medium.

Thus, if firefly $i$ detects the greater light intensity of firefly $j$ at distance $r_{ij}$, it results on a movement of $i$ towards $j$ according to
\begin{equation}
\label{coor}
x_k^i \to x_k^i + \beta(r_{ij})\left(x_k^j - x_k^i\right)  + \alpha \epsilon_k,
\end{equation}
where $x_k^i$ and $x_k^j$ are the $k$-th component of the position of the $i$-th and $j$-th fireflies, respectively, $\epsilon_k$ is a random value of the form $u - 1/2$ for $u$ from an uniform distribution (it can be substituted by a Gaussian or other distribution), and $\alpha$ is the randomization parameter, which controls the stochastic influence on the movement and is reduced recursively at every generation by a factor $\delta \in (0,1)$ on form $\alpha \to \delta \alpha$.

It is noteworthy that FA is essentially managed by three parameters: randomization parameter ($\alpha$), attractiveness ($\beta_0$), and light absorption coefficient ($\gamma$). This can be displayed as an asymptotic behavior: for instance, if $\gamma$ tends to $0$, the resulting constant attractiveness $\beta = \beta_0$ leads to a special case of PSO, and when $\gamma$ tends to $\infty$, FA becomes a random walk, parallel version of SA.
	\begin{algorithm}[H]
	\caption{Pseudocode of the firefly algorithm}
	\label{array-sum}
	\begin{algorithmic}
	\State Define objective function $f(\mathbf{x})$ where $\mathbf{x} = (x_1,\dots,x_d)^T$
    \State Assign values for $\gamma$, $\beta_0$,  $\alpha$, $\delta$ and $MaxGenerations$ (maximum number of generations)
	\State Generate initial population of fireflies $\mathbf{x}_i$ for $i=1,2,\dots,n$
	\State Determine light intensity $I_i$ at $\mathbf{x}_i$ via $f(\mathbf{x}_i)$	
	\While{$t < MaxGeneration$}
	\For{$i = 1:n$ all $n$ fireflies}
	\For{$j = 1:i$ all $n$ fireflies}
	\If{$I_i < I_j$}
	\State  Move firefly $i$ towards $j$ according (\ref{coor})
	\EndIf
	\State Vary attractiveness with distance $r$ via $\exp{\left(-\gamma r^2 \right)}$
	\State Update light intensity by evaluating new solutions
	\EndFor
	\EndFor
	\State Rank the fireflies (from lowest to highest value of the objective function) and find the current best (i.e., the last one).
    \State Update $\alpha$ by applying the reduction factor $\delta$
	\EndWhile
	\State Postprocess results and visualization
	\end{algorithmic}
	\end{algorithm}

\subsubsection{Initial parameters of the algorithm and bounding of the parametric space of the process}

As regards the initial parameters of the algorithm, and following the comments of \cite{Yang2008}, the next considerations can be made:
\begin{itemize}
	\item $\alpha$: The randomization parameter controls randomness in the movement of the fireflies. It usually takes values between 0 and 1, with a value around 0.2 being recommended.
	\item $\delta$: This parameter reduces $\alpha$ at every generation. Usually this value is chosen to be between 0.95 and 0.99.
	\item $\beta_0$: Since the attractiveness at distance $r=0$ is usually considered to be 1, $\beta_0=1$ is taken in most cases.
	\item $\gamma$: Absorption coefficient is a critical parameter that strongly influences the velocity of convergence. It simulates the environment conditions in which fireflies can detect light. In most cases, suitable values range from 1 to 10.
	\item $n$: The number of fireflies may vary in a wide range depending, for example, on the number of local maxima. Of course, having a high $n$ ensures a better coverage of the parametric space.
	\item Generations: Usually a value between 50 and 100 is considered, although a value below 50 can also provide good results.
\end{itemize}

In any case, the high degree of efficiency of FA allows us to test it with different initial values without incurring excessive computational costs.

\vskip 0.2cm

To determine the initial population of fireflies, $n$ points of the parametric space of the process must be randomly selected. To facilitate this choice, we propose bounding such space. To this end, we will use the curve reformulation made in section \ref{subsecparam} and the information provided by the sample paths:
\begin{itemize}
	\item $\lambda$: It is between 0 and 1, in order to guarantee strictly increasing curves.
	\item $\mu$: It holds $0 < \mu < \lambda^{-\sqrt{1+t_0^2}}$ for a previously established $\lambda$.
	\item $\eta$: This parameter can be bounded taking into account the asymptote of the curve verifying $k=x_{0}\left(1 + \xi(t_0)/\eta)\right)$. From that expression, and if we denote by $k_i$ the maximum value of the $i$-th sample path, the following expression holds
	\begin{equation}\label{etainterval}
	\widehat{\xi(t_0)}\left[\max\limits_{i=1,\dots,d}\left(\frac{k_i}{x_{i0}} - 1\right)\right]^{-1} < \hat{\eta} < \widehat{\xi(t_0)}\left[\min\limits_{i=1,\dots,d}\left(\frac{k_i}{x_{i0}} - 1\right)\right]^{-1},
	\end{equation}
where $x_{i0}$ is the initial value of the $i$-th sample path.
	In practice, when $t_0 \neq 0$, the algorithm must choose values for $\lambda$ and $\mu$ to finally construct an interval for $\eta$. Otherwise, in the case $t_0=0$, the interval does not depend on $\lambda$ and $\mu.$
	\item $\sigma$: Regarding parameter $\sigma$, when it has high values it leads to sample paths with great variability around the mean of the process. Thus, excessive variability in available paths would make an H1-type modeling inadvisable. Some simulations performed for several values of $\sigma$ have led us to consider that $0<\sigma<0\mbox{.}5$, so that we may have paths compatible with an H1-type growth.
\end{itemize}

\section{Applications}
\label{sec:applications}

\subsection{Application to simulated data}
We will now proceed to introduce some simulation studies. These will be aimed at validating the procedures described above as they apply to estimating the parameters of the process.

The general pattern of simulations is based on generating 30 sample paths of the H1 diffusion process in interval $[0,50]$ from (\ref{SOLEDE}). All sample paths have the same length, and $t_i=(i-1)\cdot 0.1$, $i=1,\ldots,501$ are the time instants at which observations are made. Each example has been replicated 50 times, that is, for each combination of the parameters of the model that we have considered, the simulation of the trajectories and the estimation procedure has been carried out 50 times. The final results are the average of those obtained in each replication.

\begin{figure}[H]
\centering
\includegraphics[scale=0.25]{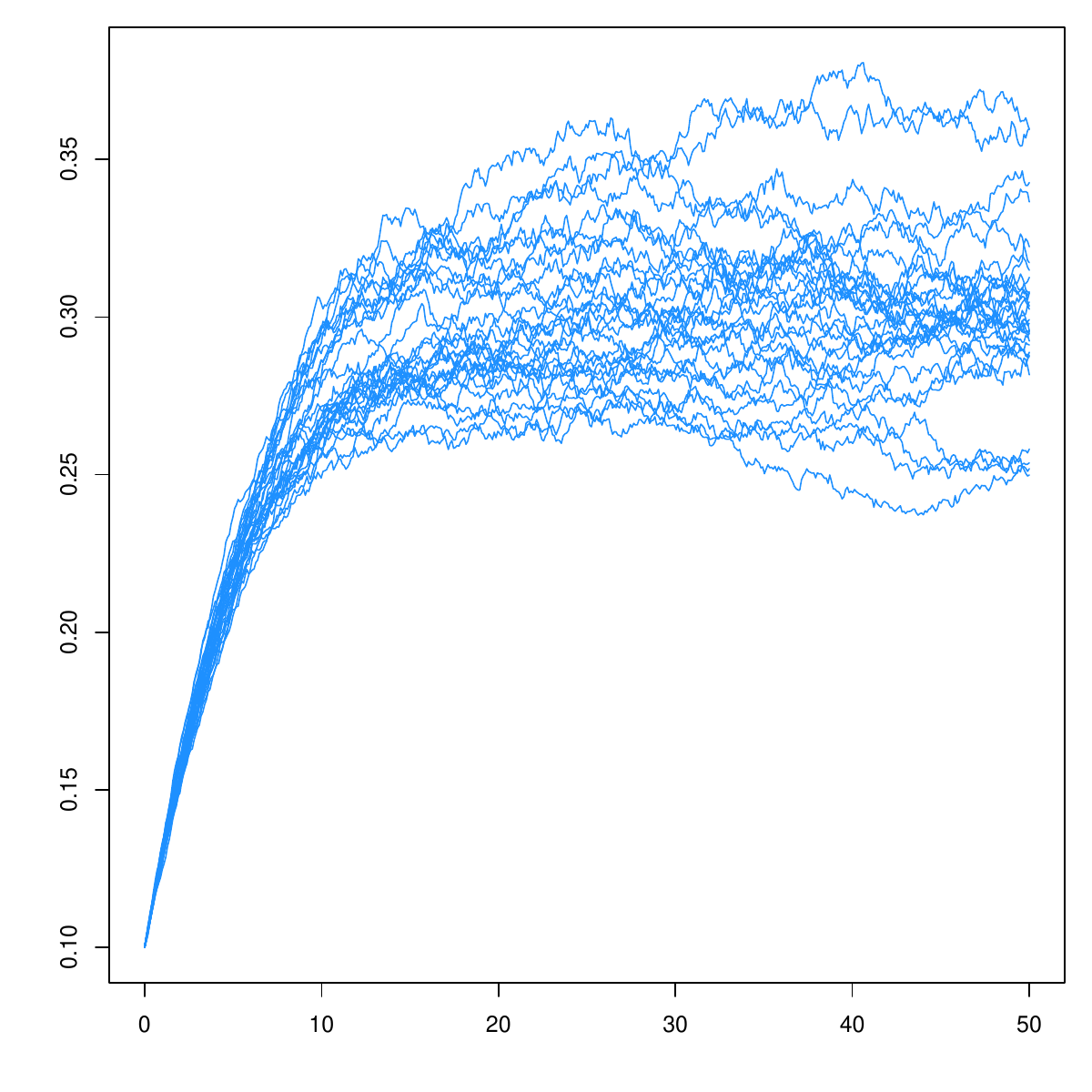}
\caption{Some simulated sample paths of the H1-type diffusion process.}
\label{fig:sims}
\end{figure}

A first simulation example has been performed by taking a degenerate initial state at $x_0=0.1$ and parameters $\eta =  0.5, \lambda = 0.8, \mu = 0.8$ and $\sigma = 0.015$ for the process. Figure \ref{fig:sims} shows the simulated paths.

We have considered $n=40$ fireflies, that is 40 $4$-uples of the parametric space, and selected each one as follows: firstly, for $\lambda$, we choose a random value (uniformly distributed) between 0 and 1  and then, for $\mu$, we take a random value in the interval $(0, 1/\lambda$) since $t_0=0$ in this case. The $4$-uple is completed by selecting a random value for $\eta$ in the interval $(0.3562706,0.5612756)$ (by virtue of \eqref{etainterval}) and for $\sigma$ in the interval $(0,0.5)$.

As regards the parameters for FA, we have made the following choices:
\begin{itemize}
\item 80 generations.
\item $\alpha = 0.2$.
\item $\delta = 0.97$.
\item $\beta_0=1$.
\item $\gamma=1$.
\end{itemize}

\begin{table}[h]
$$
\begin{array}{lccccc}
& \lambda & \mu & \eta & \sigma & f_o  \\
\hline
\mbox{Real value} & 0.8 & 0.8 & 0.5 & 0.015 & 55554.17  \\
\mbox{Estimated value} & 0.7873909  & 0.8184537 & 0.5001565 & 0.0149564 & 55554.58  \\
\mbox{Abs. relative error} & 0.0157614 & 0.0230670 & 0.0003129 & 0.0029034 & 0.0000073  \\
\hline
\end{array}
$$
\caption{Results of the firefly algorithm over simulated data.}
\label{fa}
\end{table}

Table \ref{fa} contains the real and estimated values of the parameters, as well as the value of the objective function $f_0$ at the resulting point. As a measure of the estimation error we have also included the absolute relative error, i.e., the difference in absolute value between real and estimated value, divided by the real one. From here we can observe very good results in terms of estimated values and absolute relative errors.

\begin{figure}[H]
\centering
\includegraphics[scale=.3]{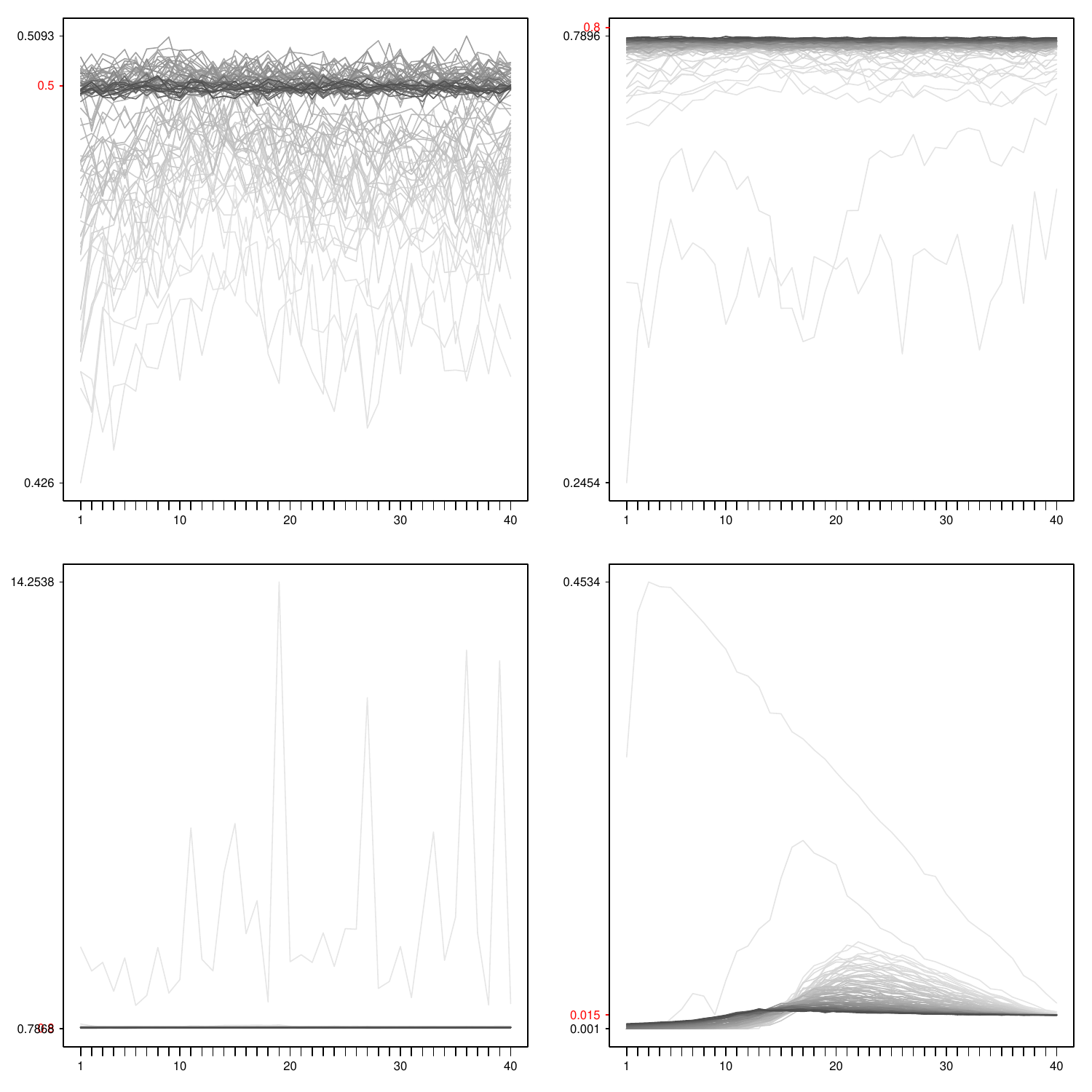}
\caption{First simulation example: evolution (from light gray to black) of 80 generations of 40 fireflies for $\eta=0.5$, $\lambda=0.8$, $\mu=0.8$ and $\sigma=0.015$ (left to right, top to bottom). Original values in red.}
\label{favary1}
\end{figure}

\begin{figure}[H]
\centering
\includegraphics[scale=.3]{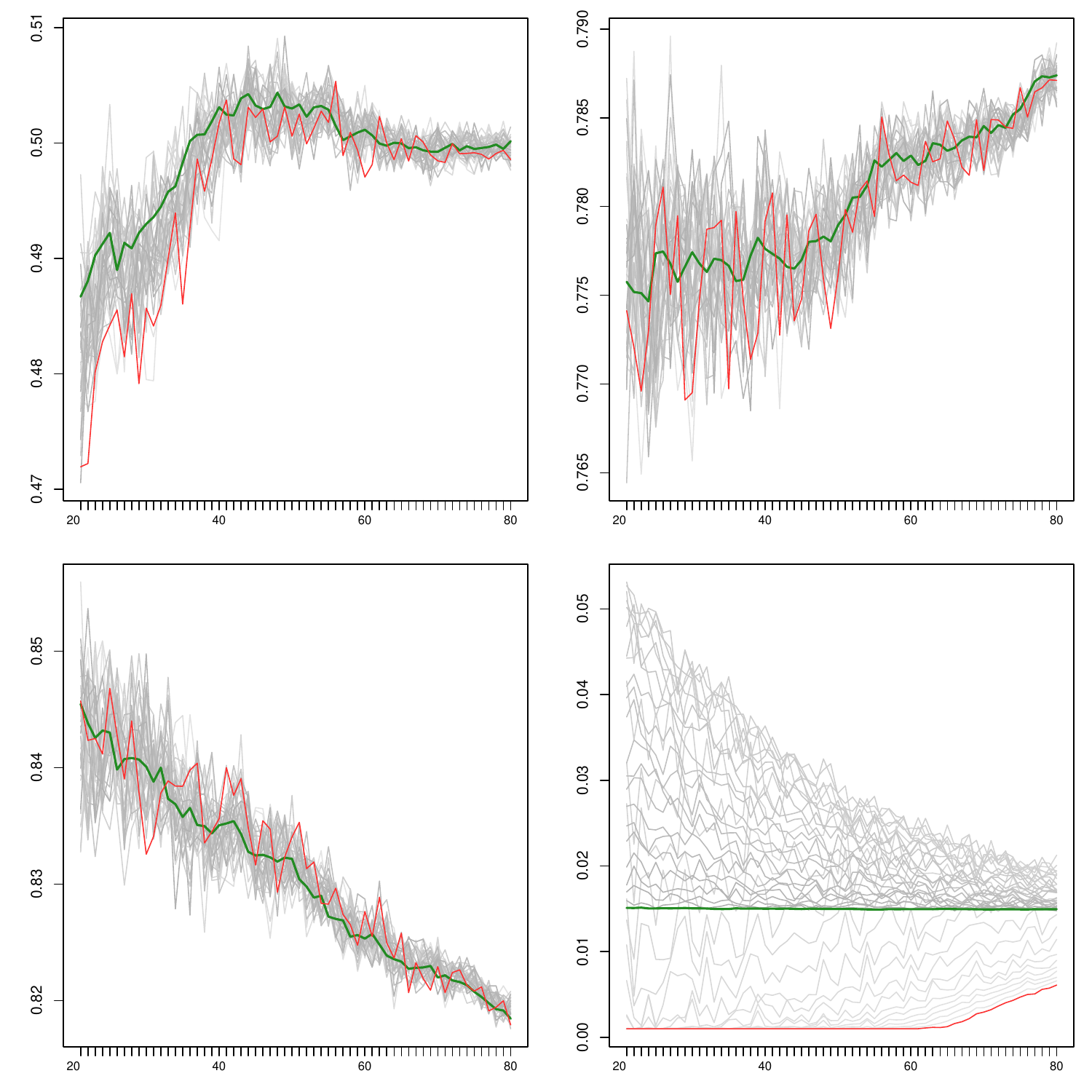}
\caption{First simulation example: evolution of fireflies over the last 60 generations for every parameter (in the same order as in the previous figure). In green, evolution of best firefly; in red, evolution of worst one.}
\label{favary2}
\end{figure}

The behavior of the algorithm over generations is shown with more detail in Figures \ref{favary1} and \ref{favary2}. For each parameter of the process, that is, for each coordinate of the fireflies, each line of Figure \ref{favary1} represents a generation (from lighter gray to dark black) showing the estimated values of the parameter for each firefly. Figure \ref{favary2} shows, for each parameter of the process, the evolution of its estimated value for each firefly over the last 60 generations. Green and red lines represent the evolution of the estimation provided by the best and the worst firefly in each generation, respectively. Remember that the best firefly is the one that provides the highest value of the objective function, whereas the worst produces the smaller one.

\begin{figure}[H]
\centering
\includegraphics[scale=.3]{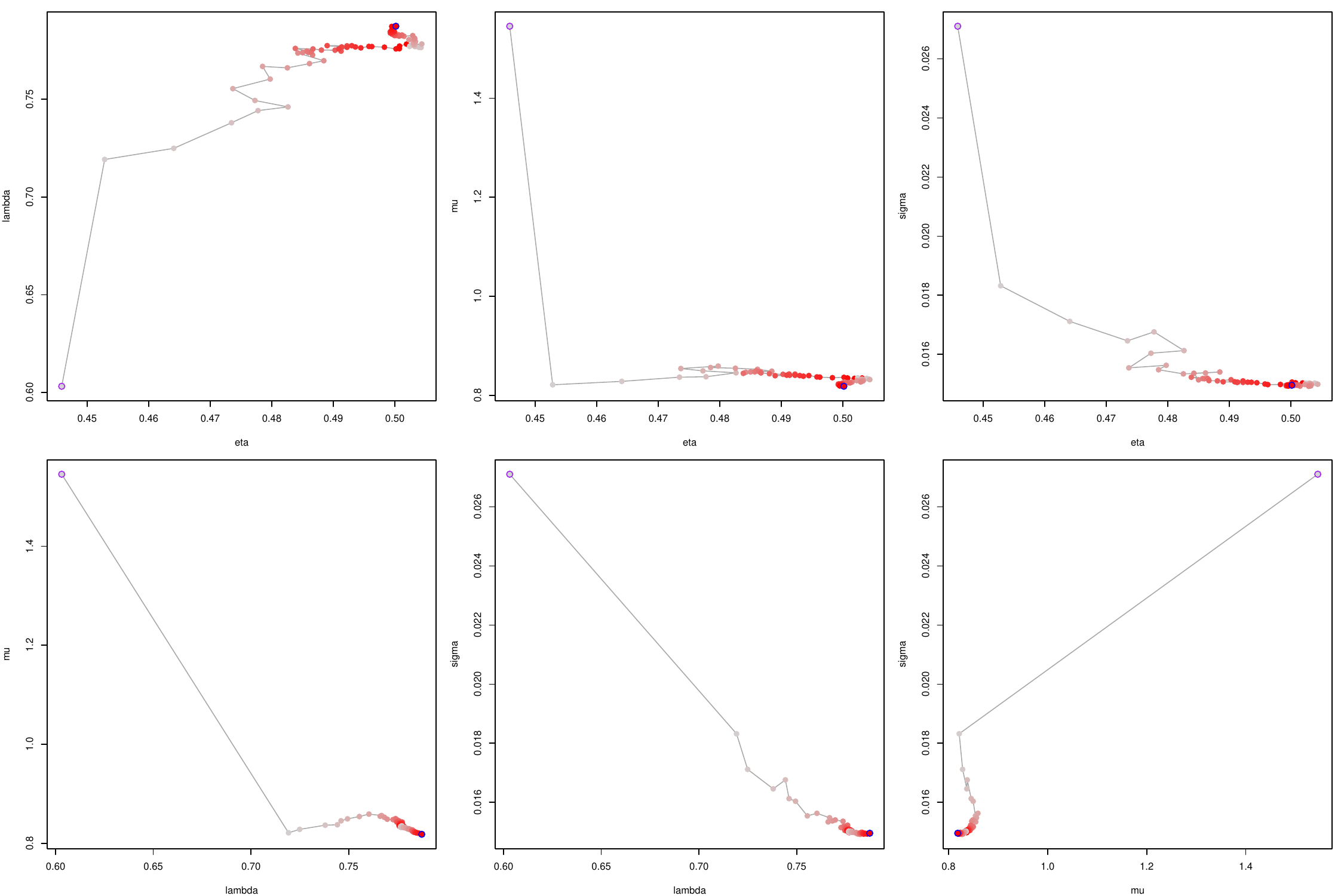}
\caption{First simulation example: for each pair of parameters, bidimensional path projection representing the evolution of the last firefly (the best) over successive generations (from light to dark red point).}
\label{simpos}
\end{figure}

Since the parametric space is 4-dimensional, it is impossible to visualize the evolution of the fireflies. For this reason, in Figure \ref{simpos} we have included graphs of the dynamics of the best firefly in each generation, for each pair of parameters.

\begin{figure}[H]
\centering
\includegraphics[scale=.3]{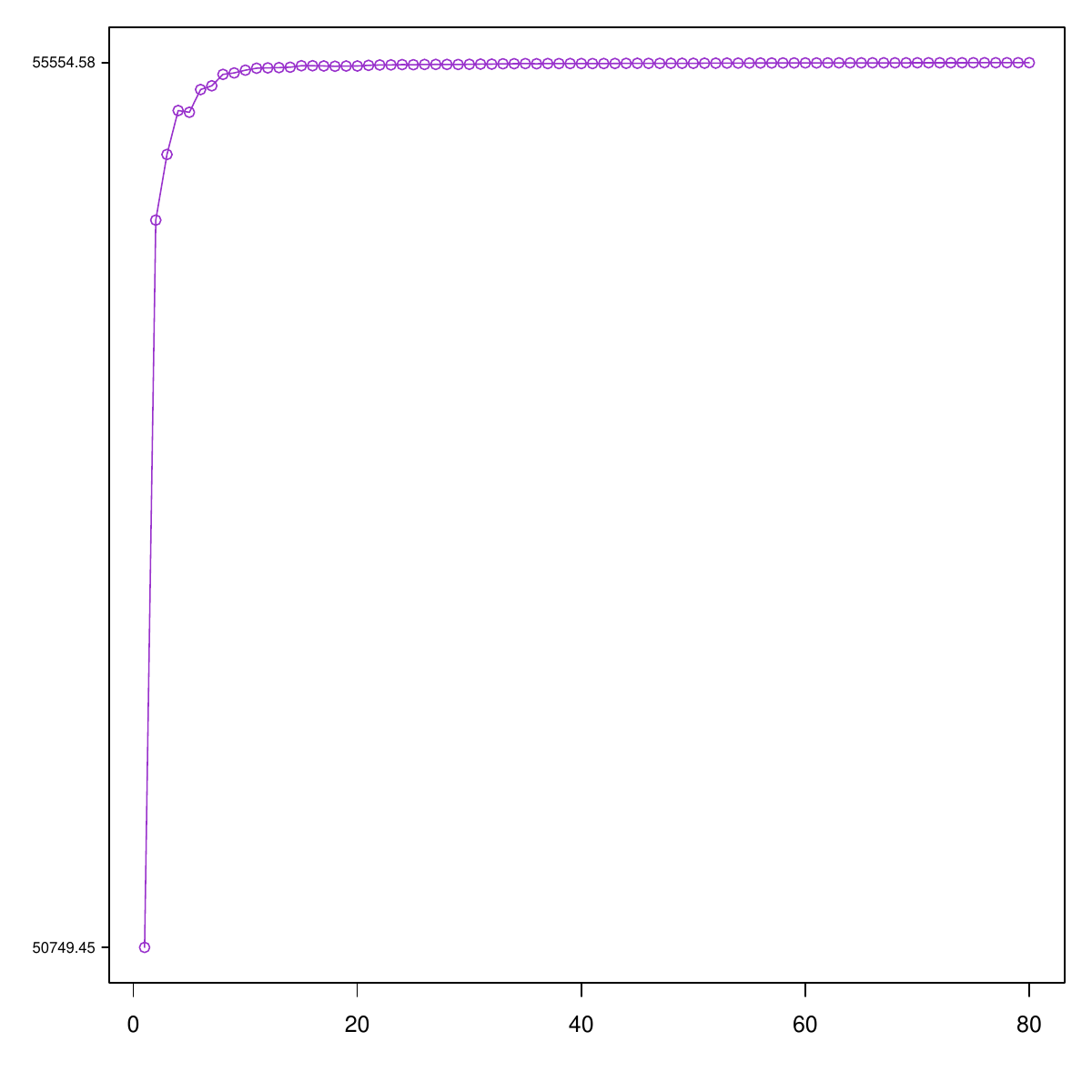}
\caption{First simulation example: value of objective function at the best firefly. Comparison between observed and fitted mean (red).}
\label{favary3}
\end{figure}

\begin{figure}[H]
\centering
\includegraphics[scale=.3]{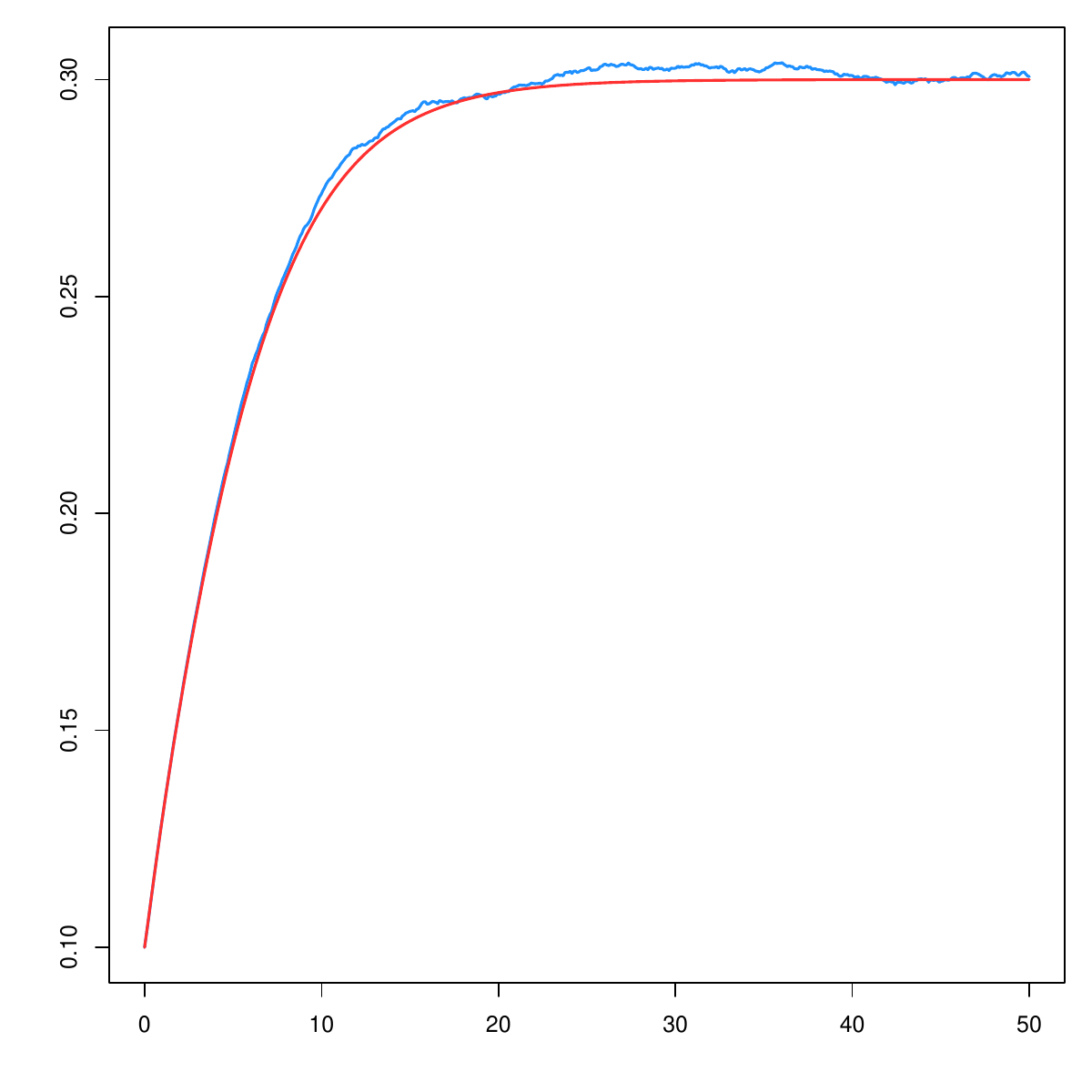}
\caption{First simulation example: comparison between observed and fitted mean (red).}
\label{favary4}
\end{figure}

Figure \ref{favary3} shows the evolution of the objective function $f_0$ for the best firefly in each generation. Obviously this graph is increasing since the best firefly of each generation improves the value of the objective function of the best of the previous generation. Finally, Figure \ref{favary4} shows the real and estimated mean function of the H1 diffusion process.

In order to observe the behavior of the algorithm for different values of the parameters, we have simulated a new dataset taking, $\eta = 0.0003,\lambda = 0.6,\mu=0.8$, $\sigma=0.025$ and $x_0 = 0.000125$. Over this data let us apply FA, varying parameters $\alpha,\gamma,\delta$ and the number of fireflies $n$, and considering 60 generations. The efficiency of the algorithm allows this type of studies to be carried out without incurring in an excessive computational cost. For example, for this simulation study, with 40 fireflies and 60 generations, the computation time has been 6 seconds for each replication, that is, 5 minutes for the total 50 replications.

Next, we analyze the behavior of absolute errors related to each pair of parameters considered when applying FA. These errors have been calculated by averaging over all other parameters.

Table \ref{Error1} shows such errors by varying $\alpha$ and $\gamma$. It can be seen that, in general, the error increases when $\gamma$ grows, regardless of the $\alpha$ value, with the most appropriate $\gamma$ values being those between 1 and 5. Conversely, for each $\gamma$, biggest errors are associated with extreme values of $\alpha$, being 0.2 and 0.4 the values for which the observed errors are smaller.

\begin{table}[!htbp]
$$
\begin{array}{cccccc}
  \multicolumn{1}{c}{} & \multicolumn{5}{c}{\gamma}\\
  \cline{2-6}
   \alpha & 1 & 5 & 10 & 20 & 35 \\
  \hline
   0.1  & 0.0199 & 0.0311 & 0.0295 & 0.0502 & 0.0454 \\
   0.2  & 0.0089 & 0.0095 & 0.0123 & 0.0220 & 0.0395 \\
   0.4  & 0.0073 & 0.0048 & 0.0111 & 0.0270 & 0.0394 \\
   0.6  & 0.0125 & 0.0115 & 0.0142 & 0.0298 & 0.0495 \\
   0.8  & 0.0112 & 0.0160 & 0.0274 & 0.0479 & 0.0642 \\
   0.9  & 0.0126 & 0.0209 & 0.0306 & 0.0459 & 0.0548 \\
\hline
\end{array}
$$
\caption{Second simulation example: absolute relative error of $\hat{f_o}$ for $\alpha$ vs. $\gamma$.}
\label{Error1}
\end{table}

Table \ref{Error2} compares absolute relative errors as a function of $\alpha$ and $\delta$. These two parameters are related since they mark the level of randomness with which the fireflies move in the space. A priori, it seems logical to think that when setting a small $\alpha$ value (low randomness), it should decrease slowly in later stages, which is associated with higher $\delta$ values. The situation should be reversed as a larger $\alpha$ initial value is selected, in which case a smaller $\delta$ value should be selected. This intuition is confirmed after observing this table. Specifically, the optimal combination is given by $\alpha=0.2$ (as suggested by \cite{Yang2008}) and $\delta=0.99$.

\begin{table}[!htbp]
$$
\begin{array}{ccccc}
  \multicolumn{1}{c}{} & \multicolumn{4}{c}{\delta}\\
  \cline{2-5}
   \alpha  & 0.9 & 0.95 & 0.97 & 0.99  \\
  \hline
          0.1  & 0.0331 & 0.0090 & 0.0052 & 0.0079 \\
          0.2  & 0.0167 & 0.0075 & 0.0047 & 0.0043 \\
          0.4  & 0.0115 & 0.0114 & 0.0114 & 0.0254 \\
          0.6  & 0.0109 & 0.0151 & 0.0166 & 0.0442 \\
          0.8  & 0.0135 & 0.0337 & 0.0426 & 0.0748 \\
          0.9  & 0.0099 & 0.0351 & 0.0461 & 0.0767 \\
\hline
\end{array}
$$
\caption{Second simulation example: absolute relative error of $\hat{f_o}$ for $\alpha$ vs. $\delta$.}
\label{Error2}
\end{table}

Table \ref{Error3} considers $\delta$ and $\gamma$ parameters. It can be seen how, in general, for fixed $\gamma$ errors grow as $\delta$ increases. Regardless of the value of $\delta$, the smallest errors are associated with $\gamma$ values between 1 and 5.

\begin{table}[!htbp]
$$
\begin{array}{cccccccc}
  \multicolumn{1}{c}{} & \multicolumn{4}{c}{\delta}\\
  \cline{2-5}
  \gamma    &  0.9 & 0.95 & 0.97 & 0.99  \\
  \hline
          1  & 0.0276 & 0.0229 & 0.0203 & 0.0228 \\
          5  & 0.0186 & 0.0161 & 0.0252 & 0.0435 \\
          10 & 0.0243 & 0.0238 & 0.0343 & 0.0489 \\
          20 & 0.0331 & 0.0364 & 0.0443 & 0.0614 \\
          35 & 0.0516 & 0.0574 & 0.0494 & 0.0635 \\
\hline
\end{array}
$$
\caption{Second simulation example: absolute relative error of $\hat{f_o}$ for $\gamma$ vs. $\delta$.}
\label{Error3}
\end{table}

Tables \ref{Error4}, \ref{Error5} and \ref{Error6} consider the combination of the number of fireflies, $n$, with the others parameters. In all cases, it is observed how the error always decreases (as seems logical) with $n$, regardless of the other parameter considered.

On the other hand, regardless of $n$, the smallest errors are associated with the central values of $\alpha$, specifically for $\alpha=0.2, 0.4$, as can be observed in Table \ref{Error4}. For fixed $n$, Table \ref{Error5} shows that the preferable $\gamma$ values are those between 1 and 10, with $\gamma = 1$ being the most recommendable. Finally, Table \ref{Error6} does not show large differences in error when $\delta$ varies.

\begin{table}[!htbp]
$$
\begin{array}{cccccc}
  \multicolumn{1}{c}{} & \multicolumn{5}{c}{n}\\
  \cline{2-6}
    \alpha & 5 & 10 & 20 & 40 & 60  \\
    \hline
          0.1  &0.1484 & 0.0387 & 0.0070 & 0.0025 & 0.0013 \\
          0.2  &0.0787 & 0.0167 & 0.0031 & 0.0014 & 0.0013 \\
          0.4  &0.0687 & 0.0150 & 0.0074 & 0.0021 & 0.0018 \\
          0.6  &0.0788 & 0.0248 & 0.0098 & 0.0054 & 0.0057 \\
          0.8  &0.0936 & 0.0331 & 0.0187 & 0.0116 & 0.0094 \\
          0.9  &0.0795 & 0.0380 & 0.0202 & 0.0126 & 0.0128 \\
\hline
\end{array}
$$
\caption{Second simulation example: absolute relative error of $\hat{f_o}$ for $\alpha$ vs. $n$.}
\label{Error4}
\end{table}

\begin{table}[!htbp]
$$
\begin{array}{cccccc}
  \multicolumn{1}{c}{} & \multicolumn{5}{c}{n}\\
  \cline{2-6}
  \gamma     & 5 & 10 & 20 & 40 & 60  \\
  \hline
          1  &0.0820 & 0.0365 & 0.0195 & 0.0075& 0.0059 \\
          5  &0.0831 & 0.0363 & 0.0234 & 0.0112& 0.0103 \\
          10 &0.0955 & 0.0399 & 0.0191 & 0.0173& 0.0126 \\
          20 &0.1584 & 0.0545 & 0.0214 & 0.0148& 0.0144 \\
          35 &0.1672 & 0.0747 & 0.0334 & 0.0185& 0.0113 \\
\hline
\end{array}
$$
\caption{Second simulation example: absolute relative error of $\hat{f_o}$ for $\gamma$ vs. $n$.}
\label{Error5}
\end{table}

\begin{table}[!htbp]
$$
\begin{array}{cccccc}
  \multicolumn{1}{c}{} & \multicolumn{5}{c}{n}\\
  \cline{2-6}
  \delta     & 5 & 10 & 20 & 40 & 60  \\
  \hline
         0.9  &  0.0911 & 0.0362 & 0.0170 & 0.0071 & 0.0038 \\
         0.95 &  0.0843 & 0.0386 & 0.0187 & 0.0080 & 0.0071 \\
         0.97 &  0.0784 & 0.0381 & 0.0259 & 0.0162 & 0.0148 \\
         0.99 &  0.1097 & 0.0568 & 0.0318 & 0.0217 & 0.0201 \\
\hline
\end{array}
$$
\caption{Second simulation example: absolute relative error of $\hat{f_o}$ for $\delta$ vs. $n$.}
\label{Error6}
\end{table}

\subsection{Application to real data}

In order to show a practical application of the process herein, we consider a real case in the context of the qPCR (Quantitative Polymerase Chain Reaction) technique. This is a method used in molecular biology to amplify DNA or RNA pieces (nucleic acids) by taking advantage of the polymerase enzyme.

Different qPCR techniques allow the simultaneous amplification and quantification of the obtained product. In particular, kinetic PCR can quantify the product from the cycle at which certain threshold of amplicon DNA is reached (the exponential phase). In order to calculate this cycle, this technique (as well as other methods) employ fluorescence monitoring, and consider the first cycle that exceeds a certain level of fluorescence. \cite{rutledge_2003} established the criteria for the mathematical development of kPCR technique, focusing on the threshold of log-fluorescence. Absolute quantification is achieved using a standard curve constructed by amplification of known amounts of target DNA and plotting the values obtained for the threshold cycle, $C_t$, against target DNA concentration.

As can be seen, the determination of the $C_t$ cycle is vital for this technique. Since we are dealing with a dynamic phenomenon, we consider fitting an H1-type diffusion process in order to model, at each instant of time (cycle in this case), the level of fluorescence. Please note that when considering a stochastic process, at each time instant we have a random variable that models the behavior of the variable under study, which allows us to take into account the random fluctuations that exist in this type of phenomena. In addition, it is possible to introduce the use of techniques associated with the study of temporal variables such as first-passage times, that is, the time at which the process verifies, for the first time, a certain property.

\begin{figure}[H]
\centering
\includegraphics[scale=0.5]{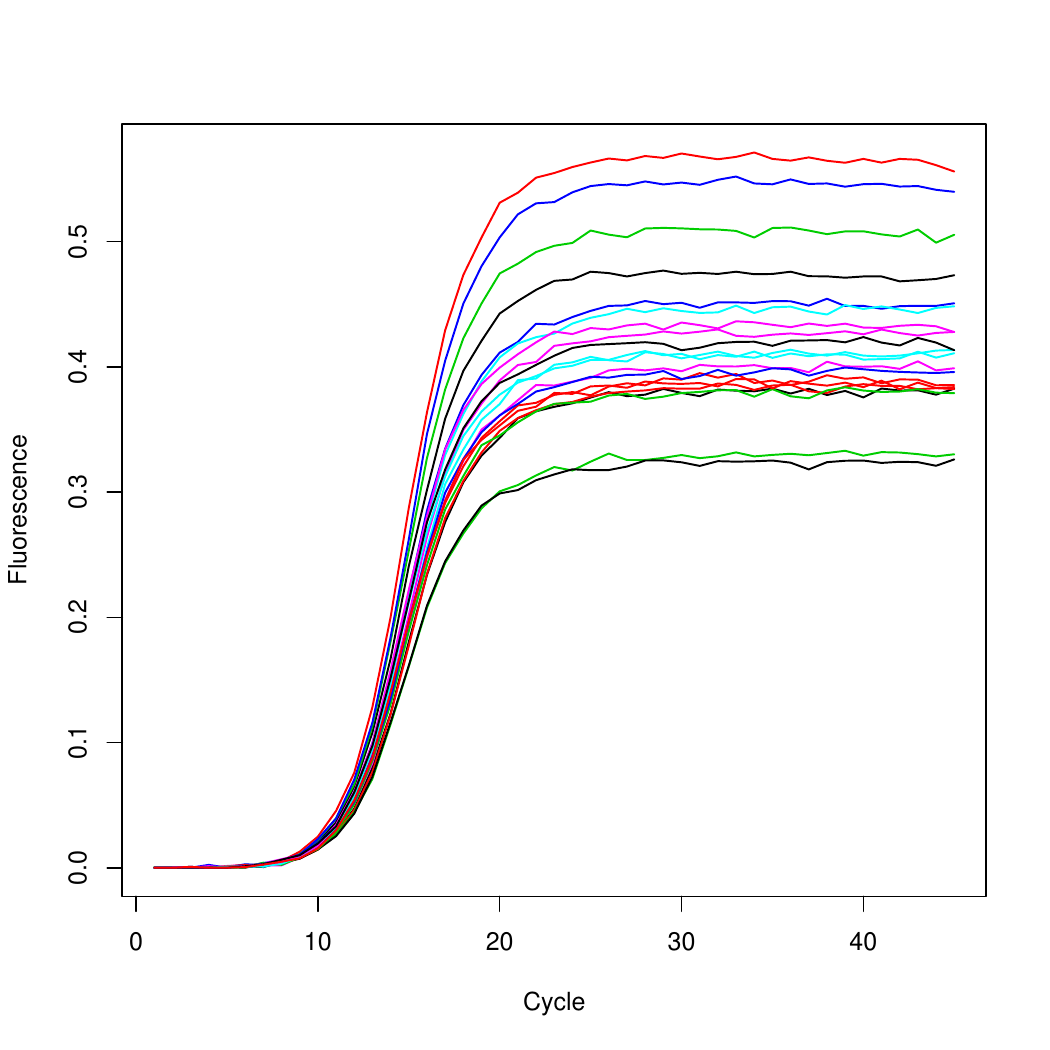}
\caption{Real application: Fluorescence versus cycle}
\label{Real_Data}
\end{figure}

The data (available as supplementary material in the paper mentioned above) corresponds to fluorescence emitted for several replicate amplifications (concretely 20) in a given concentration of DNA (as a matter of fact, in the paper the authors considered six magnitudes of target concentration, the first of which is being considered in our study) over 45 cycles (Figure \ref{Real_Data}). We have to note that, given the characteristics of the process introduced, the values considered in our study are those of fluorescence, and not its logarithm.

In order to apply FA to estimating the parameters of the process, we have considered, for the initial parameters of the algorithm, a combination of values suggested from the simulation study previously performed. Concretely, $\alpha=0.2, 0.4$, $\gamma=1,5$, $\delta=0.9 0.95, 0.97, 0.99$. In addition, we have considered $n=20$ fireflies over 110 generations. The initial time instant for the observed values is $t_1=0$, whereas for the initial value we have considered the mean of the initial values of the sample paths, being $x_1=0.000125$.

The results are summarized in Table \ref{tab:pcr}. For each combination of the initial values of the algorithm we have calculated the absolute relative errors between the observed data of the fluorescence and the corresponding estimate from the mean function of the estimated process, once the parameters of the process have been estimated. From this table we conclude that the optimal combination of the initial parameters for FA are $\alpha=0.2$, $\gamma=1$ and $\delta=0.99$.

\begin{table}[H]
\centering		
\begin{tabular}{@{\extracolsep{5pt}} ccc|ccccc}
\hline
$\alpha$ & $\gamma$ & $\delta$ & $\hat{\eta}$& $\hat{\lambda}$& $\hat{\mu}$& $\hat{\sigma}$& Error\\ \hline
0.2 &  1   & 0.9   &0.00031 &0.46869 &1.89661 &0.02499 &0.22549 \\
    &      & 0.95  &0.00029 &0.46847 &1.88574 &0.02500 &0.22092 \\
    &      & 0.97  &0.00030 &0.46274 &1.94778 &0.02500 &0.23296 \\
    &	   & 0.99  &0.00030 &0.44580 &2.27941 &0.02500 &0.17487 \\
    &  5   & 0.9   &0.00030 &0.44996 &2.05263 &0.02500 &0.26564 \\
    &      & 0.95  &0.00029 &0.46062 &1.93521 &0.02500 &0.26507 \\
    &      & 0.97  &0.00030 &0.47582 &1.80397 &0.02500 &0.22722 \\
    &	   & 0.99  &0.00032 &0.43198 &2.45740 &0.02499 &0.20623 \\
0.4 &  1   & 0.9   &0.00032 &0.46943 &1.88993 &0.02500 &0.23309 \\
    &      & 0.95  &0.00029 &0.44280 &2.26997 &0.02499 &0.22014 \\
    &      & 0.97  &0.00034 &0.44229 &2.33465 &0.02499 &0.22091 \\
    &      & 0.99  &0.00032 &0.45137 &2.21121 &0.02500 &0.19202 \\
    &  5   & 0.9   &0.00028 &0.48461 &1.66431 &0.02489 &0.29521 \\
    &      & 0.95  &0.00029 &0.42797 &2.53649 &0.02500 &0.18941 \\
    &      & 0.97  &0.00028 &0.46580 &1.86436 &0.02500 &0.29821 \\
    &	   & 0.99  &0.00031 &0.45165 &2.17721 &0.02500 &0.19100 \\
	\hline
\end{tabular}
\caption{Estimated parameter values and absolute relative errors from the application of FA to real data.}
\label{tab:pcr}
\end{table}

From these values, we have estimated the model again by increasing the number of fireflies, following the conclusions obtained in the second simulation study. Specifically, we have considered 60 fireflies. Table \ref{tab:pcr2} contains the estimates of the parameters of the process.

\begin{table}[H]
	\centering
	\begin{tabular}{*{5}{c}}	
		& $\hat\lambda$ & $\hat\mu$ & $\hat\eta$ & $\hat\sigma$  \\  \hline		
		Estimated value & 0.483548 & 1.6539 & 0.0002902 & 0.025 \\
		\hline
	\end{tabular}
	\caption{Real application: estimates of the parameters after applying FA.}
        \label{tab:pcr2}
\end{table}

\begin{figure}[H]
$$
  \begin{array}{cc}
    \includegraphics[scale=0.3]{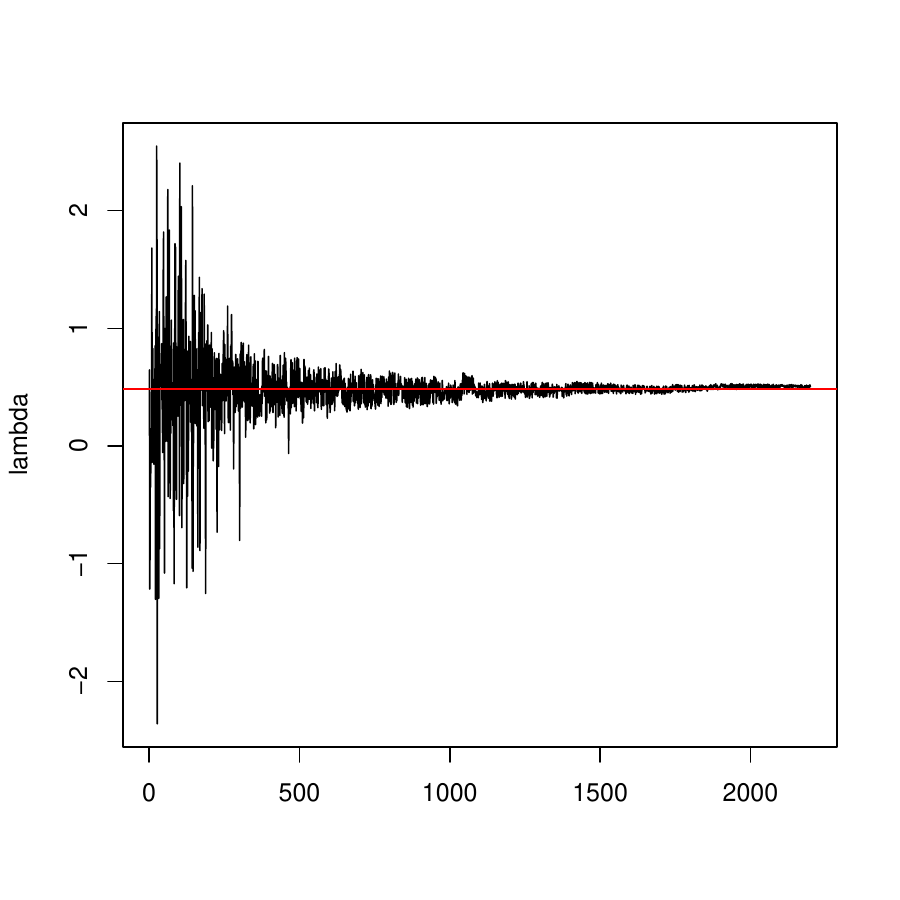} & \includegraphics[scale=0.3]{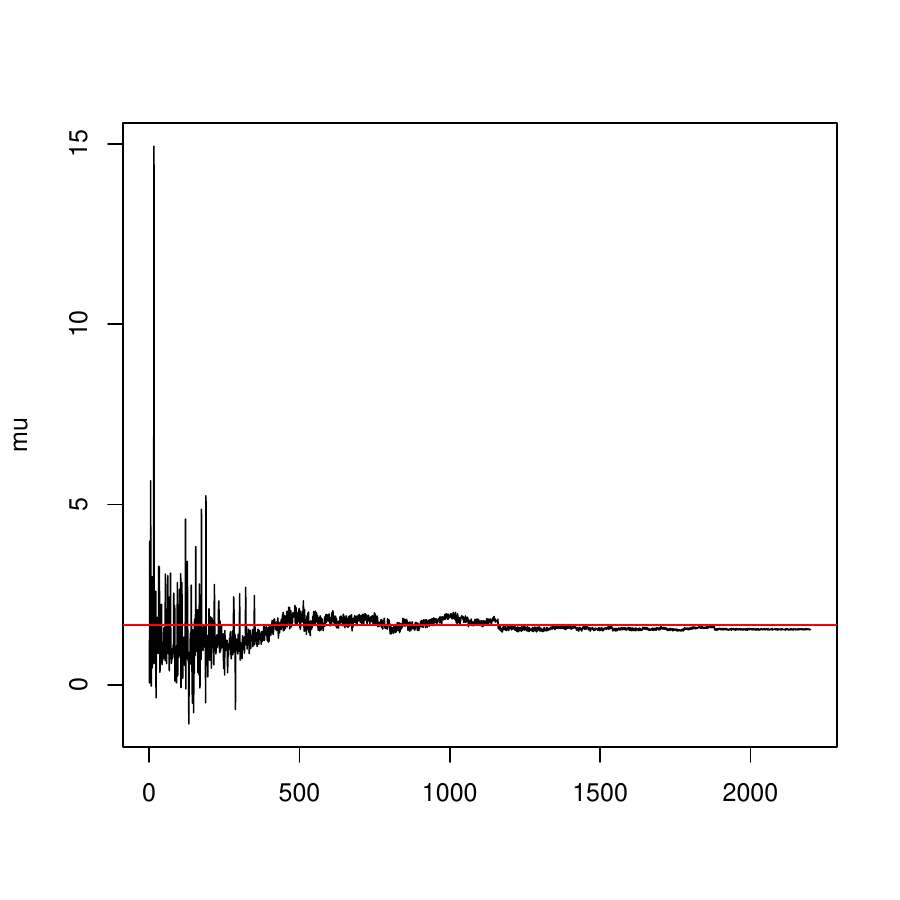}\\
	\includegraphics[scale=0.3]{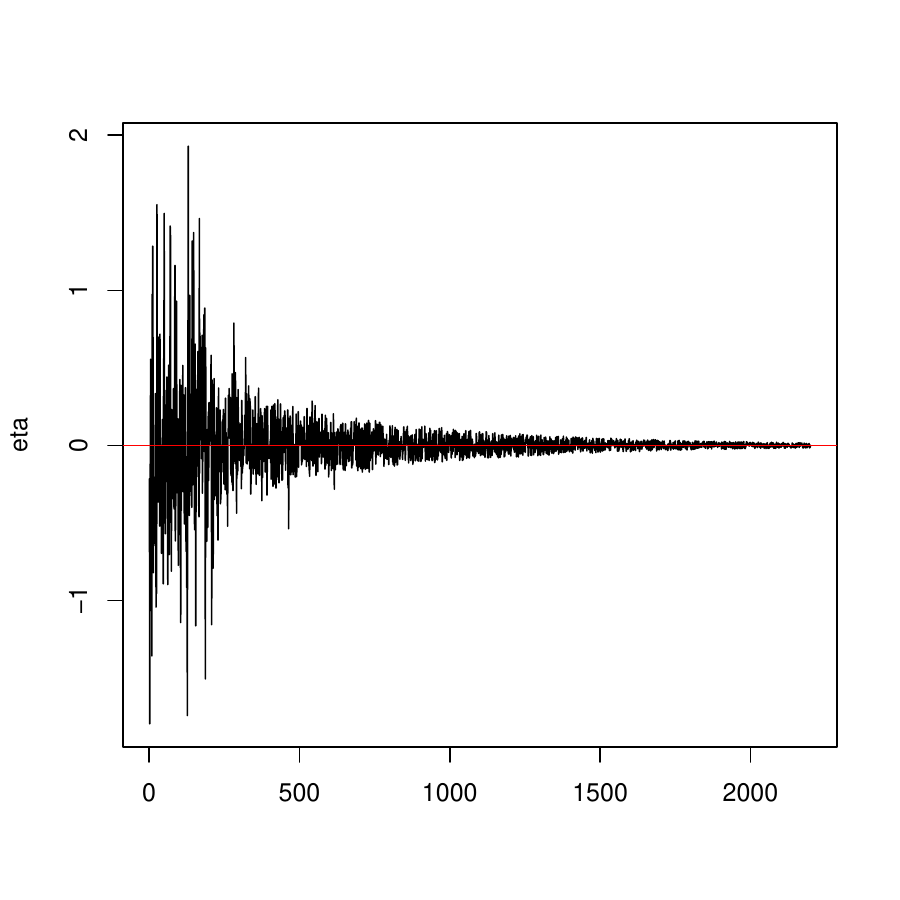} & \includegraphics[scale=0.3]{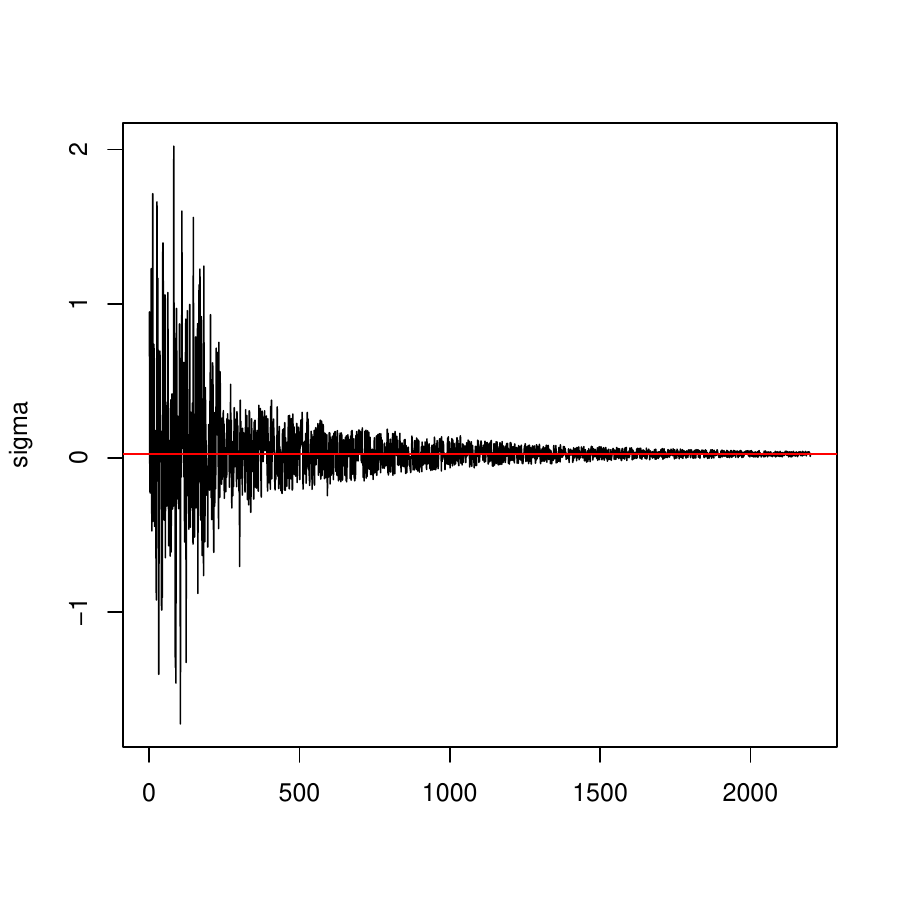}\\
  \end{array}	
  $$	
	\caption{Real application: evolution of the estimates of the parameters by applying FA.}
	\label{fig:testparam}
\end{figure}

In Figure \ref{fig:testparam} the stabilization of four estimated parameters (for the last replication) is presented.

\begin{figure}[H]
\centering
\includegraphics[scale=0.35]{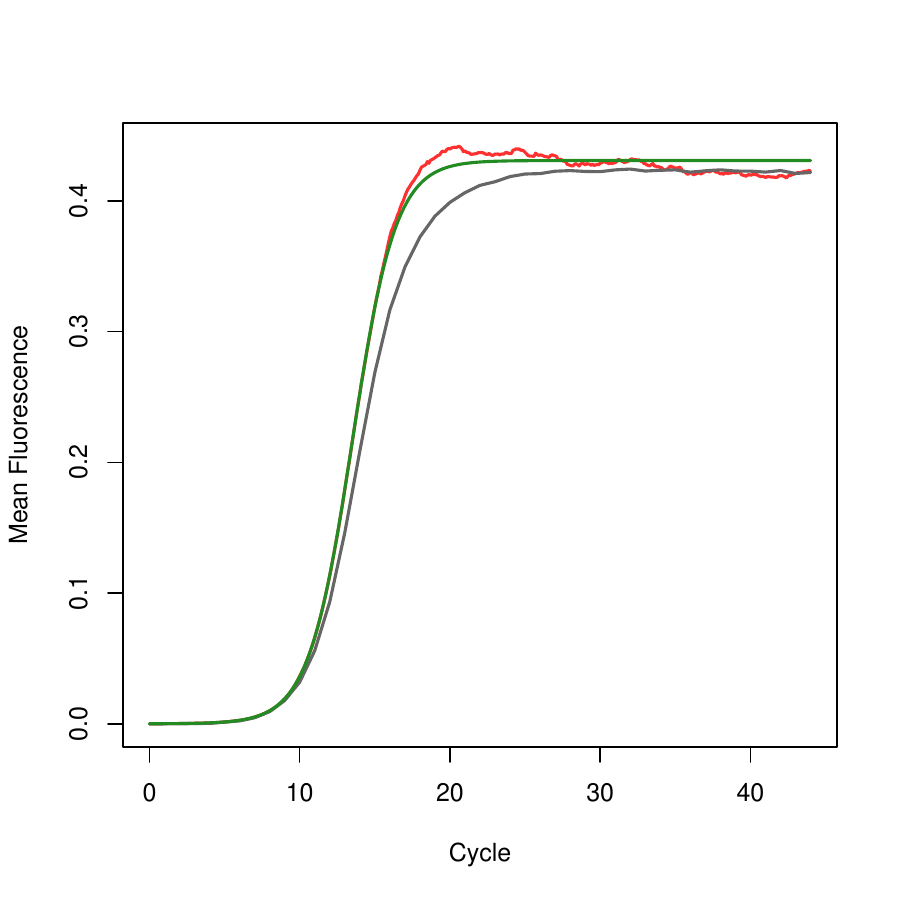}
\caption{Real application: mean of original fluorescence (grey) vs. simulated fluorescence (red) vs. fitted theoretical (green).}
\label{fig:rutfa}
\end{figure}

Once the estimates of the parameters have been obtained, we simulated the H1-type diffusion process under the conditions of the experiment. Figure \ref{fig:rutfa} shows the mean of the observed values together with the theoretical mean function and the mean of the simulated sample paths of the estimated process.

\begin{figure}[H]
\centering
\includegraphics[scale=0.35]{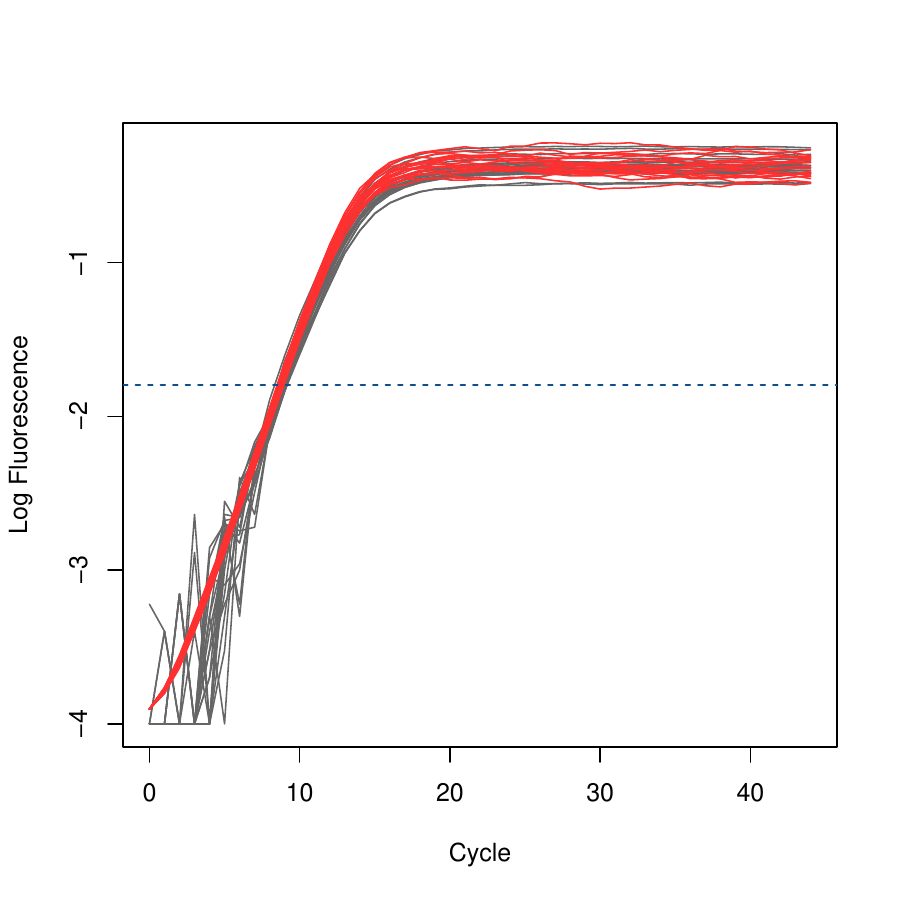}
\caption{Real application: log-fluorescence for original data (grey) and logarithm of the simulated sample paths of the simulated process (red).}
\label{fig:rutfb}
\end{figure}

With this method, we can simulate the fluorescence of the amplification procedure in different reactions, which enables us to use simulated instead of real data. Even when the simulated process differs from the original in the firsts cycles, the important question is whether we can find the cycles at which the fluorescence threshold is reached (see Figure \ref{fig:rutfb}, showing log-fluorescence values for the original data and the logarithm of the simulated sample paths of the estimated process. The horizontal line represents a fluorescence threshold). This issue can be approached using techniques associated with the study of temporal variables in the context of diffusion processes, as in the case of first-passage times.

\section{Conclusions}
The hyperbolastic type I diffusion process is obtained, showing to be advantageous over the deterministic hyperbolastic type-I curve, of wide use in many research fields. This diffusion process allows us to introduce into the model all the information coming from data, as well as the random factors which must be taken into account in order to explain different growth phenomena. Nevertheless, improvements resulting from the use of this curve are dependent on the number of parameters to estimate, and even though in this work we have reduced this quantity in one element, this complicates the development of the inference procedure. Computationally efficient methods are therefore necessary, among which metaheuristic algorithms, such as the firefly algorithm, are able to reduce computational cost.

In this work we have developed the theoretical base for the practical use of hyperbolastic type-I diffusion processes as a particular case of the lognormal process with exogenous factors, and applying the firefly algorithm in order to solve inference problems.

Simulations were performed, which show that the strategy used for bounding the parametric space behaves well, as does the firefly algorithm for different choices of its parameters. From the simulations we have obtained several combinations of the parameters of the algorithm that can be considered optimal. An example based on real data from a study about quantitative polymerase chain reaction is also developed, showing the
capability of the process for fitting the fluorescence levels associated with the amplification of amplicons of DNA.
\section*{Acknowledgements}

This work was supported in part by the Ministerio de Econom\'ia y Competitividad, Spain, under grant MTM2014-58061-P.

\section{Appendix}
\label{sec:appendix}

From the log-likelihood function (\ref{verosimilitud}), the estimates of $\eta,\lambda,\mu$ and $\sigma^2$ follow from the solution of the system of equations

\begin{align*}
&\ds \frac{\l - \T}{\vt\S} \W + \frac{\sigma^2}{2} \ds \frac{\W}{\S}  =  0 \\
&\ds \frac{\l -\T}{\vt\S}\V + \frac{\sigma^2}{2} \ds \frac{\V}{\S} =  0 \\
&\ds \frac{\l -\T}{\vt\S}\Z + \frac{\sigma^2}{2} \ds \frac{\Z}{\S} =  0  \\
&\sigma^4 \sum_{i=1}^{d} \Delta_{i1}^{in_i} + 4\sigma^2 (N-d) - 4 \ds \frac{\left(\l\right)^2}{\vt} - 4 \ds \frac{\left(\T\right)^2}{\vt} + 8 \ds \frac{\l \T}{\vt} = 0
\end{align*}
where
\begin{align*}
\S&=(\eta + \xi(t_{i,j-1}))(\eta + \xi(t_{ij}))\\
\V&=\lambda^{-1}\left[t_{i,j-1}\xi(t_{i,j-1})(\eta + \xi(t_{ij}))-t_{ij}\xi(t_{ij})(\eta + \xi(t_{i,j-1}))\right]\\
\W&=\xi(t_{ij}) - \xi(t_{i,j-1})\\
\Z&=\mu^{-1}\left[\asinh(t_{i,j-1})\xi(t_{i,j-1})(\eta + \xi(t_{ij}))-\asinh(t_{ij})\xi(t_{ij})(\eta + \xi(t_{i,j-1}))\right]
\end{align*}

Solving the last equation for $\sigma$ we obtain
\begin{equation*}
\frac{\sigma^2}{2} = U^{-\frac1{2}}\Gamma_\para^{\frac1{2}}
\end{equation*}
where
$$
\Gamma_\para = \ds \frac1{\vt}\left(\l - \T\right)^2 \ \ \mbox{and} \ \
U= 4(N-d)+ \sum_{i=1}^{d}  \Delta_{i1}^{in_i}.
$$

This expression depends only on $\eta,\lambda$ and $\mu$. Substituting in the other three equations, the following system of equations appears
\begin{equation*}
\Lambda_\mathcal{J}^\para +  U^{-\frac1{2}} \Gamma_\para^{\frac1{2}} \Psi_\mathcal{J}^\para = 0, \
\mathcal{J} = W,V,Z,
\end{equation*}
where
$$
\Lambda^\para_\mathcal{J} = \ds \frac{\l - \T}{\vt\S} \mathcal{J}_{ij}^\para \ \ \mbox{and} \ \
\Psi^\para_\mathcal{J}= \ds \frac{\mathcal{J}_{ij}^\para}{\S}, \ \ \ \mathcal{J} = W,V,Z.
$$

The solution of the above system of equations provides the maximum likelihood estimates $\hat{\eta},\hat{\lambda}$ and $\hat{\mu}$, from which $\hat{\sigma}^2 = 2U^{-\frac1{2}}\Gamma_{\hat{\eta}\hat{\lambda}\hat{\mu}}^{\frac1{2}} $.

\section*{References}

\bibliography{Biblio}

\begin{thebibliography}{33}
\expandafter\ifx\csname natexlab\endcsname\relax\def\natexlab#1{#1}\fi
\providecommand{\url}[1]{\texttt{#1}}
\providecommand{\href}[2]{#2}
\providecommand{\path}[1]{#1}
\providecommand{\DOIprefix}{doi:}
\providecommand{\ArXivprefix}{arXiv:}
\providecommand{\URLprefix}{URL: }
\providecommand{\Pubmedprefix}{pmid:}
\providecommand{\doi}[1]{\href{http://dx.doi.org/#1}{\path{#1}}}
\providecommand{\Pubmed}[1]{\href{pmid:#1}{\path{#1}}}
\providecommand{\bibinfo}[2]{#2}
\ifx\xfnm\relax \def\xfnm[#1]{\unskip,\space#1}\fi
\bibitem[{Alb et~al.(2016)Alb, Alotto, Magele, Renhart, Preis and
  Trapp}]{alb2016}
\bibinfo{author}{Alb, M.}, \bibinfo{author}{Alotto, P.},
  \bibinfo{author}{Magele, C.}, \bibinfo{author}{Renhart, W.},
  \bibinfo{author}{Preis, K.}, \bibinfo{author}{Trapp, B.},
  \bibinfo{year}{2016}.
\newblock \bibinfo{title}{Firefly algorithm for finding optimal shapes of
  electromagnetic devices}.
\newblock \bibinfo{journal}{IEEE Trans. Magn.} \bibinfo{volume}{52},
  \bibinfo{pages}{1--4}.
\newblock \DOIprefix\doi{10.1109/TMAG.2015.2483058}.
\bibitem[{Albano et~al.(2011)Albano, Giorno, Rom{\'{a}}n-Rom{\'{a}}n and
  Torres-Ruiz}]{albano_2011}
\bibinfo{author}{Albano, G.}, \bibinfo{author}{Giorno, V.},
  \bibinfo{author}{Rom{\'{a}}n-Rom{\'{a}}n, P.}, \bibinfo{author}{Torres-Ruiz,
  F.}, \bibinfo{year}{2011}.
\newblock \bibinfo{title}{Inferring the effect of therapy on tumors showing
  stochastic gompertzian growth.}
\newblock \bibinfo{journal}{J. Theor. Biol.} \bibinfo{volume}{276},
  \bibinfo{pages}{67--77}.
\newblock \DOIprefix\doi{10.1016/j.jtbi.2011.01.040}.
\bibitem[{Albano et~al.(2013)Albano, Giorno, Rom{\'{a}}n-Rom{\'{a}}n and
  Torres-Ruiz}]{Albano2013}
\bibinfo{author}{Albano, G.}, \bibinfo{author}{Giorno, V.},
  \bibinfo{author}{Rom{\'{a}}n-Rom{\'{a}}n, P.}, \bibinfo{author}{Torres-Ruiz,
  F.}, \bibinfo{year}{2013}.
\newblock \bibinfo{title}{On the effect of a therapy able to modify both the
  growth rates in a gompertz stochastic model.}
\newblock \bibinfo{journal}{Math. Biosci.} \bibinfo{volume}{245},
  \bibinfo{pages}{12--21}.
\newblock \DOIprefix\doi{10.1016/j.mbs.2013.01.001}.
\bibitem[{Barrera-Garc{\'{i}}a­ et~al.(2013)Barrera-Garc{\'{i}}a­,
  Rom{\'{a}}n-Rom{\'{a}}n and Torres-Ruiz}]{Barrera-Garcia2013a}
\bibinfo{author}{Barrera-Garc{\'{i}}a­, A.J.},
  \bibinfo{author}{Rom{\'{a}}n-Rom{\'{a}}n, P.}, \bibinfo{author}{Torres-Ruiz,
  F.}, \bibinfo{year}{2013}.
\newblock \bibinfo{title}{Fitting dynamic growth models of biological phenomena
  from sample observations through gaussian diffusion processes.}
\newblock \bibinfo{journal}{Biosyst.} \bibinfo{volume}{112},
  \bibinfo{pages}{284--291}.
\newblock \DOIprefix\doi{10.1016/j.biosystems.2012.12.007}.
\bibitem[{Capocelli and Ricciardi(1974)}]{capocelliricciardi_1974}
\bibinfo{author}{Capocelli, R.}, \bibinfo{author}{Ricciardi, L.},
  \bibinfo{year}{1974}.
\newblock \bibinfo{title}{A diffusion model for population growth in random
  environment}.
\newblock \bibinfo{journal}{Theor. Popul. Biol.} \bibinfo{volume}{5},
  \bibinfo{pages}{28--41}.
\newblock \DOIprefix\doi{10.1016/0040-5809(74)90050-1}.
\bibitem[{Deasy et~al.(2003)Deasy, Jankowski, Payne, Cao, Goff, Greenberger and
  Huard}]{deasy2003}
\bibinfo{author}{Deasy, B.}, \bibinfo{author}{Jankowski, R.},
  \bibinfo{author}{Payne, T.}, \bibinfo{author}{Cao, B.},
  \bibinfo{author}{Goff, J.}, \bibinfo{author}{Greenberger, J.},
  \bibinfo{author}{Huard, J.}, \bibinfo{year}{2003}.
\newblock \bibinfo{title}{Modeling stem cell population growth: incorporating
  terms for proliferative heterogeneity}.
\newblock \bibinfo{journal}{Stem Cells} \bibinfo{volume}{21},
  \bibinfo{pages}{536--545}.
\newblock \DOIprefix\doi{10.1634/stemcells.21-5-536}.
\bibitem[{Eby et~al.(2010)Eby, Tabatabai and Bursac}]{eby_2010}
\bibinfo{author}{Eby, W.M.}, \bibinfo{author}{Tabatabai, M.A.},
  \bibinfo{author}{Bursac, Z.}, \bibinfo{year}{2010}.
\newblock \bibinfo{title}{Hyperbolastic modeling of tumor growth with a
  combined treatment of iodoacetate and dimethylsulphoxide}.
\newblock \bibinfo{journal}{BMC Cancer} \bibinfo{volume}{10},
  \bibinfo{pages}{509}.
\newblock \DOIprefix\doi{10.1186/1471-2407-10-509}.
\bibitem[{Feller(1940)}]{Feller1940}
\bibinfo{author}{Feller, W.}, \bibinfo{year}{1940}.
\newblock \bibinfo{title}{On the logistic law of growth and its empirical
  verifications in biology}.
\newblock \bibinfo{journal}{Acta Biotheor.} \bibinfo{volume}{5},
  \bibinfo{pages}{51--66}.
\newblock \DOIprefix\doi{10.1007/BF01602862}.
\bibitem[{Fister et~al.(2013)Fister, Yang and Brest}]{fister_2013_2}
\bibinfo{author}{Fister, I.}, \bibinfo{author}{Yang, X.S.},
  \bibinfo{author}{Brest, J.}, \bibinfo{year}{2013}.
\newblock \bibinfo{title}{A comprehensive review of firefly algorithms}.
\newblock \bibinfo{journal}{Swarm Evol. Comput.} \bibinfo{volume}{13},
  \bibinfo{pages}{34--46}.
\newblock \DOIprefix\doi{doi.org/10.1016/j.swevo.2013.06.001}.
\bibitem[{Fister et~al.(2014)Fister, Yang, Brest and Fister~Jr}]{fister_2014}
\bibinfo{author}{Fister, I.}, \bibinfo{author}{Yang, X.S.},
  \bibinfo{author}{Brest, J.}, \bibinfo{author}{Fister~Jr, I.},
  \bibinfo{year}{2014}.
\newblock \bibinfo{title}{On the randomized firefly algorithm}, in:
  \bibinfo{booktitle}{Cuckoo Search and Firefly Algorithm}.
  \bibinfo{publisher}{Springer}, pp. \bibinfo{pages}{27--48}.
\bibitem[{Gandomi et~al.(2013)Gandomi, Yang, Talatahari and
  Alavi}]{gandomi_2013}
\bibinfo{author}{Gandomi, A.}, \bibinfo{author}{Yang, X.S.},
  \bibinfo{author}{Talatahari, S.}, \bibinfo{author}{Alavi, A.},
  \bibinfo{year}{2013}.
\newblock \bibinfo{title}{Firefly algorithm with chaos}.
\newblock \bibinfo{journal}{Comm. Nonlinear Sci. Numer. Simul.}
  \bibinfo{volume}{18}, \bibinfo{pages}{89--98}.
\newblock \DOIprefix\doi{10.1016/j.cnsns.2012.06.009}.
\bibitem[{Guti{\'{e}}rrez et~al.(2003)Guti{\'{e}}rrez, Rom{\'{a}}n, Romero and
  Torres}]{gutierrez_2003}
\bibinfo{author}{Guti{\'{e}}rrez, R.}, \bibinfo{author}{Rom{\'{a}}n, P.},
  \bibinfo{author}{Romero, D.}, \bibinfo{author}{Torres, F.},
  \bibinfo{year}{2003}.
\newblock \bibinfo{title}{Forecasting for the univariate lognormal diffusion
  process with exogenous factors}.
\newblock \bibinfo{journal}{Cybern. Syst.} \bibinfo{volume}{34},
  \bibinfo{pages}{709--724}.
\newblock \DOIprefix\doi{10.1080/716100279}.
\bibitem[{Guti{\'{e}}rrez et~al.(1995)Guti{\'{e}}rrez, Rom{\'{a}}n and
  Torres}]{gutierrez_1995}
\bibinfo{author}{Guti{\'{e}}rrez, R.}, \bibinfo{author}{Rom{\'{a}}n, P.},
  \bibinfo{author}{Torres, F.}, \bibinfo{year}{1995}.
\newblock \bibinfo{title}{A note on the {V}olterra integral equation for the
  first-passage-time density}.
\newblock \bibinfo{journal}{J. Appl. Prob.} \bibinfo{volume}{53},
  \bibinfo{pages}{635--648}.
\newblock \DOIprefix\doi{10.1017/S0021900200103092}.
\bibitem[{Guti{\'{e}}rrez et~al.(1999)Guti{\'{e}}rrez, Rom{\'{a}}n and
  Torres}]{gutierrez_1999}
\bibinfo{author}{Guti{\'{e}}rrez, R.}, \bibinfo{author}{Rom{\'{a}}n, P.},
  \bibinfo{author}{Torres, F.}, \bibinfo{year}{1999}.
\newblock \bibinfo{title}{Inference and first-passage-times for the lognormal
  diffusion process with exogenous factors: application to modelling in
  economics}.
\newblock \bibinfo{journal}{Appl. Stoch. Models Bus. Ind.}
  \bibinfo{volume}{15}, \bibinfo{pages}{325--332}.
\newblock
  \DOIprefix\doi{10.1002/(SICI)1526-4025(199910/12)15:4<325::AID-ASMB397>3.0.CO;2-F}.
\bibitem[{Kavousi-Fard et~al.(2014)Kavousi-Fard, Samet and
  Marzbani}]{kavousi_2014}
\bibinfo{author}{Kavousi-Fard, A.}, \bibinfo{author}{Samet, H.},
  \bibinfo{author}{Marzbani, F.}, \bibinfo{year}{2014}.
\newblock \bibinfo{title}{A new hybrid modified firefly algorithm and support
  vector regression model for accurate short term load forecasting}.
\newblock \bibinfo{journal}{Expert Syst. Appl.} \bibinfo{volume}{41},
  \bibinfo{pages}{6047--6056}.
\newblock \DOIprefix\doi{10.1016/j.eswa.2014.03.053}.
\bibitem[{Niknam et~al.(2012)Niknam, Azizipanah-Abarghooee and
  Roosta}]{niknam_2012}
\bibinfo{author}{Niknam, T.}, \bibinfo{author}{Azizipanah-Abarghooee, R.},
  \bibinfo{author}{Roosta, A.}, \bibinfo{year}{2012}.
\newblock \bibinfo{title}{Reserve constrained dynamic economic dispatch: a new
  fast self-adaptive modified firefly algorithm}.
\newblock \bibinfo{journal}{IEEE Syst. J.} \bibinfo{volume}{6},
  \bibinfo{pages}{635--646}.
\newblock \DOIprefix\doi{10.1109/JSYST.2012.2189976}.
\bibitem[{Oksendal(2003)}]{oksendal_2003}
\bibinfo{author}{Oksendal, B.}, \bibinfo{year}{2003}.
\newblock \bibinfo{title}{Stochastic differential equations: an introduction
  with applications}.
\newblock \bibinfo{edition}{6th} ed., \bibinfo{publisher}{Springer-Verlag},
  \bibinfo{address}{New York}.
\bibitem[{Rom{\'{a}}n-Rom{\'{a}}n et~al.(2012)Rom{\'{a}}n-Rom{\'{a}}n, Romero,
  Rubio and Torres-Ruiz}]{roman_2012}
\bibinfo{author}{Rom{\'{a}}n-Rom{\'{a}}n, P.}, \bibinfo{author}{Romero, D.},
  \bibinfo{author}{Rubio, M.}, \bibinfo{author}{Torres-Ruiz, F.},
  \bibinfo{year}{2012}.
\newblock \bibinfo{title}{Estimating the parameters of a {G}ompertz-type
  diffusion process by means of simulated annealing}.
\newblock \bibinfo{journal}{App. Math. Comput.} \bibinfo{volume}{218},
  \bibinfo{pages}{5121--5131}.
\newblock \DOIprefix\doi{10.1016/j.amc.2011.10.077}.
\bibitem[{Rom{\'{a}}n-Rom{\'{a}}n et~al.(2010)Rom{\'{a}}n-Rom{\'{a}}n, Romero
  and Torres-Ruiz}]{roman_2010}
\bibinfo{author}{Rom{\'{a}}n-Rom{\'{a}}n, P.}, \bibinfo{author}{Romero, D.},
  \bibinfo{author}{Torres-Ruiz, F.}, \bibinfo{year}{2010}.
\newblock \bibinfo{title}{A diffusion process to model generalized von
  {B}ertalanffy growth patterns: Fitting to real data}.
\newblock \bibinfo{journal}{J. Theor. Biol.} \bibinfo{volume}{263},
  \bibinfo{pages}{59--69}.
\newblock \DOIprefix\doi{10.1016/j.jtbi.2009.12.009}.
\bibitem[{Rom{\'a}n-Rom{\'a}n and Torres-Ruiz(2014)}]{roman_2014}
\bibinfo{author}{Rom{\'a}n-Rom{\'a}n, P.}, \bibinfo{author}{Torres-Ruiz, F.},
  \bibinfo{year}{2014}.
\newblock \bibinfo{title}{Forecasting fruit size and caliber by means of
  diffusion processes. application to ``valencia late'' oranges}.
\newblock \bibinfo{journal}{J. Agric. Biol. Envir. S.} \bibinfo{volume}{19},
  \bibinfo{pages}{292--313}.
\newblock \DOIprefix\doi{10.1007/s13253-014-0172-3}.
\bibitem[{Rom{\'{a}}n-Rom{\'{a}}n and Torres-Ruiz(2015)}]{roman_2015}
\bibinfo{author}{Rom{\'{a}}n-Rom{\'{a}}n, P.}, \bibinfo{author}{Torres-Ruiz,
  F.}, \bibinfo{year}{2015}.
\newblock \bibinfo{title}{A stochastic model related to the {R}ichards-type
  growth curve. estimation by means of simulated annealing and variable
  neighborhood search}.
\newblock \bibinfo{journal}{App. Math. Comput.} \bibinfo{volume}{266},
  \bibinfo{pages}{579 -- 598}.
\newblock \DOIprefix\doi{10.1016/j.amc.2015.05.096}.
\bibitem[{Russo et~al.(2009)Russo, Baldi, Parisi, Magnifico, Mariani and
  Cataudella}]{russo_2009}
\bibinfo{author}{Russo, T.}, \bibinfo{author}{Baldi, P.},
  \bibinfo{author}{Parisi, A.}, \bibinfo{author}{Magnifico, G.},
  \bibinfo{author}{Mariani, S.}, \bibinfo{author}{Cataudella, S.},
  \bibinfo{year}{2009}.
\newblock \bibinfo{title}{Levy processes and stochastic von {B}ertalanffy
  models of growth, with application to fish population analysis}.
\newblock \bibinfo{journal}{J. Theor. Biol.} \bibinfo{volume}{258},
  \bibinfo{pages}{521--529}.
\newblock \DOIprefix\doi{10.1016/j.jtbi.2009.01.033}.
\bibitem[{Rutledge and Cote(2003)}]{rutledge_2003}
\bibinfo{author}{Rutledge, R.}, \bibinfo{author}{Cote, C.},
  \bibinfo{year}{2003}.
\newblock \bibinfo{title}{Mathematics of quantitative kinetic pcr and the
  application of standard curves}.
\newblock \bibinfo{journal}{Nucleic Acids Res.} \bibinfo{volume}{31},
  \bibinfo{pages}{e93}.
\newblock \DOIprefix\doi{10.1093/nar/gng093}.
\bibitem[{Schurz(2007)}]{schurz_2007}
\bibinfo{author}{Schurz, H.}, \bibinfo{year}{2007}.
\newblock \bibinfo{title}{Modeling, analysis and discretizations of stochastic
  logistic equations}.
\newblock \bibinfo{journal}{Int. J. Numer. Anal. Mod.} \bibinfo{volume}{4},
  \bibinfo{pages}{178--197}.
\bibitem[{Tabatabai et~al.(2012)Tabatabai, Eby and Bursac}]{tabatabai_2012}
\bibinfo{author}{Tabatabai, M.}, \bibinfo{author}{Eby, W.},
  \bibinfo{author}{Bursac, Z.}, \bibinfo{year}{2012}.
\newblock \bibinfo{title}{Oscillabolastic model, a new model for oscillatory
  dynamics, applied to the analysis of {H}es1 gene expression and {E}hrlich
  ascites tumor growth}.
\newblock \bibinfo{journal}{J. Biomed. Inform.} \bibinfo{volume}{45},
  \bibinfo{pages}{401--407}.
\newblock \DOIprefix\doi{10.1016/j.jbi.2011.11.016}.
\bibitem[{Tabatabai et~al.(2005)Tabatabai, Williams and
  Bursac}]{tabatabai_2005}
\bibinfo{author}{Tabatabai, M.}, \bibinfo{author}{Williams, D.},
  \bibinfo{author}{Bursac, Z.}, \bibinfo{year}{2005}.
\newblock \bibinfo{title}{Hyperbolastic growth models: theory and application}.
\newblock \bibinfo{journal}{Theor. Biol. Med. Model.} \bibinfo{volume}{2},
  \bibinfo{pages}{14}.
\newblock \DOIprefix\doi{10.1186/1742-4682-2-14}.
\bibitem[{Tabatabai et~al.(2011)Tabatabai, Bursac, Eby and
  Singh}]{tabatabai_2011}
\bibinfo{author}{Tabatabai, M.A.}, \bibinfo{author}{Bursac, Z.},
  \bibinfo{author}{Eby, W.M.}, \bibinfo{author}{Singh, K.P.},
  \bibinfo{year}{2011}.
\newblock \bibinfo{title}{Mathematical modeling of stem cell proliferation}.
\newblock \bibinfo{journal}{Med. Biol. Eng. Comput.} \bibinfo{volume}{49},
  \bibinfo{pages}{253--262}.
\newblock \DOIprefix\doi{10.1007/s11517-010-0686-y}.
\bibitem[{Tabatabai et~al.(2013)Tabatabai, Eby, Singh and
  Bae}]{tabatabai_2013_x}
\bibinfo{author}{Tabatabai, M.A.}, \bibinfo{author}{Eby, W.M.},
  \bibinfo{author}{Singh, K.P.}, \bibinfo{author}{Bae, S.},
  \bibinfo{year}{2013}.
\newblock \bibinfo{title}{T model of growth and its application in systems of
  tumor-immune dynamics}.
\newblock \bibinfo{journal}{Math. Biosci. Eng.} \bibinfo{volume}{10},
  \bibinfo{pages}{925--938}.
\newblock \DOIprefix\doi{10.3934/mbe.2013.10.925}.
\bibitem[{Tsoularis and Wallace(2002)}]{tsoularis_2002}
\bibinfo{author}{Tsoularis, A.}, \bibinfo{author}{Wallace, J.},
  \bibinfo{year}{2002}.
\newblock \bibinfo{title}{Analysis of logistic growth models}.
\newblock \bibinfo{journal}{Math. Biosci.} \bibinfo{volume}{179},
  \bibinfo{pages}{21--55}.
\newblock \DOIprefix\doi{10.1016/S0025-5564(02)00096-2}.
\bibitem[{Tuckwell and Koziol(1987)}]{tuckwell_1987}
\bibinfo{author}{Tuckwell, H.C.}, \bibinfo{author}{Koziol, J.A.},
  \bibinfo{year}{1987}.
\newblock \bibinfo{title}{Logistic population growth under random dispersal}.
\newblock \bibinfo{journal}{B. Math. Biol.} \bibinfo{volume}{49},
  \bibinfo{pages}{495 -- 506}.
\newblock \DOIprefix\doi{10.1016/S0092-8240(87)80010-1}.
\bibitem[{Yang(2008)}]{Yang2008}
\bibinfo{author}{Yang, X.S.}, \bibinfo{year}{2008}.
\newblock \bibinfo{title}{Nature-Inspired Metaheuristic Algorithms}.
\newblock \bibinfo{publisher}{Luniver Press}.
\bibitem[{Yang and He(2013)}]{yang_2013_y}
\bibinfo{author}{Yang, X.S.}, \bibinfo{author}{He, X.}, \bibinfo{year}{2013}.
\newblock \bibinfo{title}{Firefly algorithm: recent advances and applications}.
\newblock \bibinfo{journal}{Internat. J. Swarm Intell.} \bibinfo{volume}{1},
  \bibinfo{pages}{36--50}.
\newblock \DOIprefix\doi{10.1504/IJSI.2013.055801}.
\bibitem[{Zhang et~al.(2016)Zhang, Liu, Yang and Dai}]{zang_2016}
\bibinfo{author}{Zhang, L.}, \bibinfo{author}{Liu, L.}, \bibinfo{author}{Yang,
  X.S.}, \bibinfo{author}{Dai, Y.}, \bibinfo{year}{2016}.
\newblock \bibinfo{title}{A novel hybrid firefly algorithm for global
  optimization}.
\newblock \bibinfo{journal}{PLoS ONE} \bibinfo{volume}{11},
  \bibinfo{pages}{1--17}.
\newblock \DOIprefix\doi{10.1371/journal.pone.0163230}.

\end{thebibliography}

\end{document}